\documentclass[12pt]{article}

\usepackage{epsf,amsfonts,hyperref}
\newcommand{\newsection}{    
\setcounter{equation}{0}
\section}
\renewcommand{\appendix}[1]{
    \addtocounter{section}{1}
    \setcounter{equation}{0}
    \renewcommand{\thesection}{\Alph{section}}
    \newsection*{Appendix \thesection\protect\indent #1}
    \addcontentsline{toc}{section}{Appendix \thesection\ \ \ #1}
}
\newcommand\encadremath[1]{\vbox{\hrule\hbox{\vrule\kern8pt
\vbox{\kern8pt \hbox{$\displaystyle #1$}\kern8pt}
\kern8pt\vrule}\hrule}}
\def\enca#1{\vbox{\hrule\hbox{
\vrule\kern8pt\vbox{\kern8pt \hbox{$\displaystyle #1$}
\kern8pt} \kern8pt\vrule}\hrule}}

\newcommand\figureframex[3]{
\begin{figure}[bth]
\hrule\hbox{\vrule\kern8pt
\vbox{\kern8pt \vbox{
\begin{center}
{\mbox{\epsfxsize=#1.truecm\epsfbox{#2}}}
\end{center}
\caption{#3}
}\kern8pt}
\kern8pt\vrule}\hrule
\end{figure}
}
\newcommand\figureframey[3]{
\begin{figure}[bth]
\hrule\hbox{\vrule\kern8pt
\vbox{\kern8pt \vbox{
\begin{center}
{\mbox{\epsfysize=#1.truecm\epsfbox{#2}}}
\end{center}
\caption{#3}
}\kern8pt}
\kern8pt\vrule}\hrule
\end{figure}
}

\newcommand{\beq}{\begin{equation}}
\newcommand{\eeq}{\end{equation}}
\newcommand{\bea}{\begin{eqnarray}}
\newcommand{\eea}{\end{eqnarray}}

\renewcommand{\and}{{\qquad {\rm and} \qquad}}

\newcommand{\virg}{{\qquad , \qquad}}

 \newcommand{\Tr}{{\,\rm Tr}\:}
\newcommand{\tr}{{\,\rm tr}\:}

\newcommand{\Res}{\mathop{\,\rm Res\,}}

\newcommand{\td}[1]{{\tilde{#1}}}

\renewcommand{\l}{\lambda}

\newcommand{\ee}[1]{{{\rm e}^{#1}}}

\renewcommand{\d}{{{\partial}}}
\newcommand{\D}{{{\hbox{d}}}}

\newcommand{\Pint}{{\int\kern -1.em -\kern-.25em}}

\newcommand{\ssq}[1]{{\sqrt{\sigma({#1})}}}

\renewcommand{\l}{\lambda}

\renewcommand{\ssq}[1]{{\sqrt{\sigma({#1})}}}
\newcommand{\ovl}{\overline}

\begin{document}
\sloppy


\pagestyle{empty}
\hfill SPT-04/086
\addtolength{\baselineskip}{0.20\baselineskip}
\begin{center}
\vspace{26pt}
{\large \bf {Topological expansion for the 1-hermitian matrix model correlation functions.
}}
\newline
\vspace{26pt}

{\sl B.\ Eynard}\hspace*{0.05cm}\footnote{ E-mail: eynard@spht.saclay.cea.fr }\\
\vspace{6pt}
Service de Physique Th\'{e}orique de Saclay,\\
F-91191 Gif-sur-Yvette Cedex, France.\\
\end{center}

\vspace{20pt}
\begin{center}
{\bf Abstract}:
\end{center}

We rewrite the loop equations of the hermitian matrix model, in a way which involves no derivative with respect to the potential,
we compute all the correlation functions, to all orders in the topological $1/N^2$ expansion, as residues on an hyperelliptical curve.
Those residues, can be represented as Feynmann graphs of a cubic field theory on the curve.

%





\vspace{26pt}
\pagestyle{plain}
\setcounter{page}{1}


\newsection{Introduction}

We consider the formal hermitian matrix integral:
\beq\label{Zdef}
Z= \int_{H_N} \D{M}\, \ee{-N \tr V(M)}
\eeq
where $M$ is a $N\times N$ hermitian matrix,  $\D{M}$ is the product of Lebesgue measures of all real components of $M$.
$V(x)$ is a polynomial of degree $d+1\geq 2$ called the potential.
$Z$ is called the partition function.

Our goal is to compute the large $N$ limit, as well as the full $1/N^2$ expansion,
 of the following formal expectation values ($<>$ is the average computed with the probability measure ${1\over Z}\, \ee{-N\tr V(M)}\, dM$):
\beq\label{defWkbar}
\overline{W}_k(x_1,\dots,x_k) := N^{k-2} \left< \tr{1\over x_1-M}\tr{1\over x_2-M}\dots \tr{1\over x_k-M} \right>_{\rm c}
\eeq
\beq\label{defWk}
W_k(x_1,\dots,x_k) := \mathop{\lim}_{N\to\infty} \overline{W}_k(x_1,\dots,x_k)
\eeq
where the subscript $c$ means connected part or cumulant.

When $Z$ is considered as a formal generating function, it is well known \cite{ZJDFG} that
the correlation function $\overline{W}_k(x_1,\dots,x_k)$ has a $1/N^2$ expansion, also called topological expansion, noted:
\beq\label{defWkgenush}
\overline{W}_k(x_1,\dots,x_k) := \sum_{h=0}^\infty N^{-2h} W^{(h)}_k(x_1,\dots,x_k)
\eeq

Let us emphasize that in general $Z$ is not a convergent integral,
the partition function as well as the $\overline{W}_k$'s are to be understood as
formal series in the coefficients of the potential, see \cite{ZJDFG} for details.

In that formal sense, the expectation value of a product of $k$ traces is the combinatoric generating function
for enumerating discrete surfaces with $k$ holes, and the variables $x_1,\dots,x_k$ are fugacities for
the lengths of the $k$ boundaries \cite{GinspargGQ2D, Matrixsurf, BIPZ, grossGQ2D}.
It is well known \cite{ZJDFG} that the power of $N$ associated to each discrete surface is its Euler characteristic $\chi$.
For a genus $h$ connected surface with $k$ holes, we have $\chi=2-k-2h$.
This is why the quantity \ref{defWkbar} has a large $N$ limit.
The large $N$ limit \ref{defWk} $W_k(x_1,\dots,x_k)=W^{(0)}_k(x_1,\dots,x_k)$
is therefore the generating function of genus zero discrete surfaces with $k$ boundaries,
and each $W^{(h)}_k(x_1,\dots,x_k)$ is the generating function of genus $h$ discrete surfaces with $k$ boundaries.

\medskip
The problem of computing the $\ovl{W}_k$'s has been addressed many times, for various applications to physics and mathematics.
Indeed, the correlation functions of eigenvalues (and thus of traces of powers) of a random matrix
have a universal behavior which (this is what universality means)
is observed in many physical phenomena, ranging from solid state physics (quantum chaos, mesoscopic conductors, see \cite{Mehta, guhr} )
to high energy physics (nuclear physics \cite{Mehta}, Quantum chromodynamics \cite{verbar}, string theory \cite{DV}), and in mathematics (distribution of Riemann zeta's zeroes \cite{BI}).

In the 90's, random matrices were extensively studied in the context of quantum gravity (see \cite{ZJDFG}), which is nothing but statistical physics on
random discretized surfaces, i.e.  the combinatorial problem of enumerating discrete surfaces of given topology, as described above.

Quantum gravity is also deeply related to conformal field theory (CFT), when one takes a ``double scaling limit'' where very large discrete surfaces are dominant,
in other words, CFT is the limit of continuous surfaces.
Depending on the limit chosen, and on the coefficients of the potential $V(x)$, one may reach different double scaling limits, which are in relationships with
the $(p,q=2)$ minimal models in CFT. All the critical exponents of such surfaces are given by KPZ's formula \cite{KPZ}.

It is thus expected, that in appropriate double scaling limits, expectation values of the form \ref{defWkbar}, can be computed from a quantum field theory,
namely Liouville's theory.

Here, without taking any double scaling limit,we will find a quantum-field-theory-like Feynmann expansion for the $\ovl{W}_k(x_1,\dots,x_k)$.

\medskip

The $W_k$'s have been computed in the literature by various methods. A formula of Dyson \cite{dyson}
 gives the $\overline{W}_k$'s for finite $N$ in terms of orthogonal polynomials, but is
not very convenient for large $N$ limit calculations, and is not convenient for the formal model.
The method of loop equations \cite{ZJDFG, staudacher, akeman, ACKM, eynardchain, eynm2m} gives recursion relations between the $\overline{W}_k$'s, which simplify in the large $N$ limit.
The loop equations have been known for a while, and give a very effective algorithm for computing explicitly the
$W_k$'s (see \cite{akeman,akemanambjorn, ACKM}).
The method developed by \cite{akeman,akemanambjorn} for computing the $W_k$'s, consists in computing $W_1$ and then
obtain the $W_k$'s by taking iterated derivatives with respect to the potentials (loop insertion method).
This method has two drawbacks: first in order to find $W_{k+1}$, one should know $W_k$ for all potentials (in particular one must take infinite degree potentials);
second, before computing $W_k$, one has to compute $W_1,W_2,\dots,W_{k-1}$, i.e. it has not been found how to integrate the recursion formulae of \cite{akemanambjorn}.

Here we consider new loop equations, which allow to find recursion relations between the $W_k$'s, without taking any derivatives with respect to the potential
(we may work with fixed potential).
Moreover, the recursion relations for the $W_k^{(h)}$'s obtained in this paper can be integrated:
the $k^{\rm th}$-loop function to order $N^{-2h}$ is a $k$-legs, $h$-loops Feynmann graph of a $\phi^3$ theory living on an hyperelliptical curve.

\bigskip

{\noindent \bf Outline:}

- in section \ref{defnotations} we introduce the notations.

- in section \ref{algeom} we introduce some basic tools of algebraic geometry.

- in section \ref{loopeqlargeN} we write and solve the loop equations to large $N$ leading order, i.e. we compute the $W_k$'s.

- in section \ref{loopeqhighergenus} we write and solve the loop equations recursively to each order in $1/N^2$, i.e. we compute the $W^{(h)}_k$'s.

- in section \ref{onecut} we do explicitly the computation in the one-cut case.

- in section \ref{concl} we conclude by presenting perspectives of applications to other matrix models (2 matrix model).


\newsection{Definitions and notations}\label{defnotations}

From now on, we assume that $V'(x)$ is monic of degree $d\geq 1$.

\subsection{Loop functions}

For $k\geq 1$, we define (the subscript $c$ means connected part or cumulant):
\beq
\overline{W}_k(x_1,\dots,x_k) := N^{k-2} \left< \tr{1\over x_1-M}\tr{1\over x_2-M}\dots \tr{1\over x_k-M} \right>_{\rm c}
\eeq
\beq
\overline{U}_k(x_1;x_2,\dots,x_k) := N^{k-2} \left< \tr{V'_1(x_1)-V'_1(M_1)\over x_1-M}\tr{1\over x_2-M}\dots \tr{1\over x_k-M} \right>_{\rm c}
\eeq
and their large $N$ limits:
\beq
W_k(x_1,\dots,x_k) := W_k^{(0)}(x_1,\dots,x_k) :=\mathop{\lim}_{N\to\infty} \overline{W}_k(x_1,\dots,x_k)
\eeq
\beq
U_k(x_1;x_2,\dots,x_k) := U^{(0)}_k(x_1;x_2,\dots,x_k) :=\mathop{\lim}_{N\to\infty} \overline{U}_k(x_1;x_2,\dots,x_k)
\eeq
as well as their formal $1/N^2$ expansions ($k\geq 1$, $h\geq 0$):
\beq
\overline{W}_k(x_1,\dots,x_k) := \sum_{h=0}^\infty N^{-2h}\, W^{(h)}_k(x_1,\dots,x_k)
\eeq
\beq
\overline{U}_k(x_1;x_2,\dots,x_k) := \sum_{h=0}^\infty N^{-2h}\, U^{(h)}_k(x_1;x_2,\dots,x_k)
\eeq

Notice $U_1^{(0)}$ is a monic polynomials of degree $d-1$,
and as soon as $k+h\geq 2$, $U^{(h)}_k$ is a polynomial of
degree at most $d-2$ in $x_1$.
We have:
\beq
{U}^{(h)}_k(x_1;x_2,\dots,x_k)
= \mathop{\rm Pol}_{x_1\to\infty} V'_1(x_1)\, {W}^{(h)}_k(x_1,\dots,x_k)
\eeq

The functions $\overline{W}_k$ are called loop-functions, because they are generating functions
for discrete surfaces with $k$ boundaries, i.e. $k$ loops.

\subsection{Filling fractions}

If the integral \ref{Zdef} were to be considered as a convergent integral, the $1/N^2$ expansion would exist only
in the so-called one-cut case (see \cite{BDE, dkmvz}).
Here \ref{Zdef} is considered as a formal power series, by its expansion in the vicinity of a minimum of the potential
$\tr V(M)$.
The potential $V(x)$ has in general $d=\deg V'$ extrema, and thus, the potential $\Tr V(M)$ can have extrema indexed by the number of
eigenvalues of $M$ lying in the vicinity of each extrema of $V(x)$.
The formal perturbative expansion around such local extrema cannot change the number of eigenvalues near each extrema.
The fractions of eigenvalues near each extrema of $V$ are called filling fractions, and are thus moduli characterizing the
vacuum near which the perturbative formal expansion is computed.
The filling fractions play an important role in recent applications of random matrix models to string theory \cite{DV}.

The filling fractions are denoted:
\beq
\epsilon_1,\epsilon_2,\dots,\epsilon_{s}
\,\, , \qquad
\sum_{j=1}^s \epsilon_j = 1
\eeq

It is well known \cite{akeman, ZJDFG} (and we recover it below) that the function $W_1(x)$ is solution of an algebraic equation,
it has $s$ cuts $[a_{2j-1},a_{2j}]$, $j=1,\dots,s$, which correspond to the location of eigenvalues in the large $N$ limit.
The condition that the filling fractions are given can be written:
\beq\label{acycleepsilonintro}
\forall j=1,\dots,s
\,\, , \qquad
{1\over 2i\pi} \oint_{[a_{2j-1},a_{2j}]}\, W_1(x)\, \D{x} = \epsilon_j
\eeq
where the contour surrounds the segment $[a_{2j-1},a_{2j}]$ in the trigonometric direction.

\medskip
Let us for a moment, use the method of \cite{akemanambjorn, akeman} for finding the filling fraction conditions for other loop functions.

If $V(x)=\sum_k t_k x^k$, from \cite{akemanambjorn} we introduce the loop insertion operator\footnote{The loop insertion operator
$\sum_{k=1}^\infty\, {1\over x^{k+1}}\,{\partial \over \partial t_k}$ is a formal notation which makes sense order
by order in the $1/x$ expansion, and eq \ref{Wkacyclevanishintro} is perfectly rigorously proven.}:
\beq\label{defddV}
{\partial\over \partial V(x)}:=-\sum_{k=1}^\infty\, {1\over x^{k+1}}\,{\partial \over \partial t_k}
\eeq
we then have \cite{akeman}:
\beq\label{dWkdV}
W_{k+1}(x_1,\dots,x_k,x_{k+1}) = {\partial \over \partial V(x_{k+1})}\, W_{k}(x_1,\dots,x_{k})
\eeq
and thus, since the filling fractions are given parameters independent of $V$ and $N$, we must have for all $k\geq 1$, $h\geq 0$, $h+k>1$:
\beq\label{Wkacyclevanishintro}
\forall j=1,\dots,s
\,\, , \qquad
 \oint_{[a_{2j-1},a_{2j}]} W^{(h)}_k(x_1,x_2,\dots,x_k)\, \D{x_1} = 0
\eeq
From the same argument, since we assume that there is no eigenvalue elsewhere in the complex plane, we can write,
for any $m$ in the complex plane, away from the cuts:
\beq\label{W1mcirclevanish}
 \oint_{m} W_1(x_1)\, \D{x_1} = 0
\eeq
(where the contour integral is a small circle around $m$)
and thus:
\beq\label{Wkmcirclevanish}
 \oint_{m} W^{(h)}_k(x_1,x_2,\dots,x_k)\, \D{x_1} = 0
\eeq

\newsection{The one-loop function and algebraic geometry}\label{algeom}

\subsection{The one--loop function}

It is well known \cite{ZJDFG, akeman} (and it is re-derived below) that the one loop function is algebraic:
\beq\label{W1}
W_1(x) = {1\over 2} \left( V'(x)-M(x)\ssq{x}\right)
\eeq
\beq\label{U1}
U_1(x) = {1\over 4}\left( V'^2(x)-M^2(x)\sigma(x) \right)
\eeq
where $M$ and $\sigma$ are monic polynomials (remember $V'(x)$ is monic), determined by:
\beq\label{largexasymp}
W_1(x) \mathop\sim_{x\to \infty} {1\over x}
\eeq
and by \ref{acycleepsilonintro}, which can be rewritten as follows:
let $a_1,\dots, a_{2s}$ be the zeroes of $\sigma$:
\beq
\sigma(x) = \prod_{i=1}^{2s} (x-a_i)
\eeq
we must have:
\beq\label{zerobcycles}
\forall\, j\in[1,s-1],\qquad
\int_{a_{2j-1}}^{a_{2j}} M(x)\ssq{x}\, \D{x} = 2i\pi \epsilon_j
\eeq
For a given $s\in[1,d]$, the equations \ref{largexasymp} and \ref{zerobcycles}
give a finite number of solutions for $M$ and $\sigma$.
Let us assume that we have chosen one of them.

\subsection{More notations}

For convenience we introduce $m_1,\dots, m_{d-s}$ the zeroes of $M$:
\beq
M(x) = \prod_{i=1}^{d-s} (x-m_i)
\eeq
We also define for $k\geq 1$, $h\geq 0$, and $h+k>1$:
\beq\label{defFkh}
F_k^{(h)}(x_1,\dots,x_k) := \left(2^k W_k(x_1,\dots,x_k)+{2\delta_{k,2}\delta_{h,0}\over (x_1-x_2)^2} \right)\, \prod_{i=1}^k \ssq{x_i}
\eeq
and:
\beq\label{defFk}
F_k(x_1,\dots,x_k):=F_k^{(0)}(x_1,\dots,x_k)
\eeq
It is well known that the $F_k$'s and $F_k^{(h)}$'s are rational functions of all their arguments (see \cite{akemanambjorn,akeman,ACKM}).

Another useful notation is in terms of multi-linear differential forms:
\beq\label{defGk}
G_k(x_1,\dots,x_k) := W_k(x_1,\dots,x_k)\, dx_1\dots dx_k
\eeq
and for higher orders:
\beq\label{defGkh}
G^{(h)}_k(x_1,\dots,x_k) := W^{(h)}_k(x_1,\dots,x_k)\, dx_1\dots dx_k
\eeq
It is well known that they are all multi-linear differentials defined on an hyperelliptical surface.
All of them, except $G_1$ and $G_2$, have poles only at the branch-points (i.e. the zeroes of $\sigma$),
and have vanishing contour integrals around the cuts.

All this is re-derived below.

\subsection{Hyperelliptical surfaces}

We need to introduce some basic notions of algebraic geometry \cite{Farkas, Fay}.

Equation \ref{W1} defines an hyperelliptical surface of genus $s-1$.
Let $y=V'(x)-2W_1(x)$, we have:
\beq\label{defy}
y^2 = M^2(x)\sigma(x)
\eeq
That equation defines a Riemann surface with two sheets (corresponding to the two determinations of the square root).
In other word, for each $x$, there are two values of $y(x)$.

Let us define the physical sheet as the sheet where:
\beq
x^{-s}\ssq{x}\mathop{\sim}_{x\to\infty} +1
\eeq
and the second sheet as the one where:
\beq
x^{-s}\ssq{x}\mathop{\sim}_{x\to\infty} -1
\eeq
If $x$ is a point in the physical sheet, let us note $\overline{x}$ the point corresponding
to the same $x$ in the second sheet.
By definition, we have:
\beq
\ssq{\ovl{x}}=-\ssq{x}
\virg
M(\ovl{x})=M(x)
\virg
y(\ovl{x})=-y(x)
\virg
d\ovl{x}=dx
\eeq

The branch points $a_i$ are the points where the two sheets meet, they are such that:
\beq
\forall i=1,\dots,2s \, , \quad \ovl{a}_i=a_i
\eeq

Near a branch point $a_i$, the surface is better parameterized by the local coordinate:
\beq
\tau_i(x):=\sqrt{x-a_i}=-\tau_i(\ovl{x})
\eeq
i.e.
\beq
x=a_i+\tau_i^2
\virg
dx = 2\tau_i d\tau_i
\eeq
In particular, the differential $dx$ has a (simple) zero at $x=a_i$.

\medskip
{\bf \noindent Holomorphic differentials}:
Let $L(x)$ be any polynomial of degree $\leq s-2$.
Since $\ssq{x}$ has a simple zero at $x=a_i$, the differential $L(x){dx\over \ssq{x}}$ has no singularity on the whole surface
(neither near the branch points, nor at $\infty$), it is thus called a holomorphic differential.
One has the following classical theorem:
there exist a unique set of $s-1$ polynomials of degree $s-2$, which we note $L_j(x)$,
such that:
\beq\label{defholomform}
\forall\, l,j\in[1,s-1]^2,\qquad
\oint_{[a_{2l-1},a_{2l}]} {L_j(x)\over \ssq{x}}\, \D{x} = 2i\pi\,\delta_{l,j}
\eeq
The differentials ${L_j(x)\over \ssq{x}}dx$ are called the normalized holomorphic differentials.
Notice that the $L_j$'s form a basis of degree $\leq s-2$ polynomials.
For any polynomial $P(x)$ such that $\deg P\leq s-2$, we have:
\beq\label{decompholomform}
P(x) = \sum_{j=1}^{s-1} \left({1\over 2i\pi}\,\oint_{[a_{2j-1},a_{2j}]} {P(x')\over \ssq{x'}}\, \D{x'}\right)\, L_j(x)
\eeq

Notice that on the $s^{\rm th}$ cut, we have:
\beq
\forall\, j\in[1,s-1],\qquad
\oint_{[a_{2s-1},a_{2s}]} {L_j(x)\over \ssq{x}}\, \D{x} = - 2i\pi
\eeq

We define:
\beq
L_s(x):=0
\eeq
so that \ref{decompholomform} holds also with the sum on $j$ running from 1 to $s$.

\medskip
{\bf \noindent Normalized differential of the third kind}:
For any $x'$ on the curve, there exists a unique meromorphic differential, noted $dS(x,x')$,
which has only two simple poles in $x$, located at $x=x'$ and $x=\ovl{x'}$, and such that:
\beq\label{defdS}
\left\{\begin{array}{l}
\displaystyle dS(x,x')\mathop\sim_{x\to x'} {dx\over x-x'}+{\rm finite} \cr
\displaystyle dS(x,x')\mathop\sim_{x\to \ovl{x'}} -{dx\over x-x'}+{\rm finite} \cr
\displaystyle \forall j=1,\dots,s-1\, , \quad \oint_{[a_{2j-1},a_{2j}]} dS(x,x') = 0 \cr
\end{array}\right.
\eeq
Notice that $dS(x,x')$ is a meromorphic differential in the variable $x$, and a multi-valued function of the variable $x'$.

It is easy to check that we have the following expression:
\beq\label{CjLambda}
dS(x,x')=
{\ssq{x'}\over \ssq{x}}\,\left(
{1\over x-x'}-\sum_{j=1}^{s-1} C_j(x')L_j(x)
\right)\, dx
\eeq
where
\beq\label{defCj}
C_j(x'):={1\over 2i\pi}\,\oint_{[a_{2j-1},a_{2j}]} {\D{x}\over \ssq{x}}\,{1\over x-x'}
\eeq
In this formula, it is assumed that $x'$ lies outside the contours $[a_{2j-1},a_{2j}]$.
One has to be careful when $x'$ approaches some branch point $a_j$.
When $x'$ lies inside the contour around $[a_{2j-1},a_{2j}]$, then one has:
\beq
C_l(x')+{\delta_{l,j}\over \ssq{x'}}={1\over 2i\pi}\,\oint_{[a_{2l-1},a_{2l}]} {\D{x}\over \ssq{x}}{1\over x-x'}
\eeq
which is analytical in $x'$ when $x'$ approaches $a_{2j-1}$ or $a_{2j}$.

For $i=1,\dots, s$, we define:
\bea\label{defdSi}
dS_{2i-1}(x,x')
&:=& dS_{2i}(x,x') := dS(x,x')-{L_i(x)\over \ssq{x}} \cr
&=& {\ssq{x'}\over \ssq{x}}\,\left(
{1\over x-x'}-{L_i(x)\over \ssq{x'}}-\sum_{j=1}^{s-1} C_j(x')L_j(x)
\right)\, dx
\eea
which is a one-form in $x$, with poles at $x=x'$ and $x=\ovl{x'}$, and which is analytical in $x'$ when $x'$ is close to $a_{2i-1}$ or $a_{2i}$.

\medskip
{\bf \noindent Bergmann kernel}:
For any $x'$ on the curve, there exists a unique bilinear differential, noted $B(x,x')$,
called the Bergmann kernel,
which has only one double pole in $x$, located at $x=x'$ (in particular no pole at $x=\ovl{x'}$), with no residue, and such that:
\beq\label{defBergmann}
\left\{\begin{array}{l}
\displaystyle B(x,x')\mathop\sim_{x\to x'} {dx \, dx'\over (x-x')^2}+{\rm finite} \cr
\displaystyle \forall j=1,\dots,s-1\, , \quad \oint_{x\in [a_{2j-1},a_{2j}]} B(x,x') = 0 \cr
\end{array}\right.
\eeq
It is easy to check that $B(x,x')=B(x',x)$ and:
\bea\label{BergmanCjdS}
B(x,x')& = & {1\over 2\ssq{x}}\,dx\, dx'\,{\d\over \d x'}
\left({\ssq{x}+\ssq{x'}\over x-x'} - \sum_{j=1}^{s-1} C_j(x')L_j(x)\ssq{x'}\right) \cr
& = & {1\over 2}\, dx'\,{\d\over \d x'}
\left({dx\over x-x'}+dS(x,x')\right)
\eea
It can be written:
\beq\label{BergmanQ}
B(x,x') = {dx dx'\over 2(x-x')^2} + {Q(x,x')dx dx'\over 4(x-x')^2\ssq{x}\ssq{x'}}
\eeq
where $Q(x,x')$ is a symmetric polynomial in $x$ and $x'$, of degree at most $s$,
such that
 $Q(x,x)=2\sigma(x)$ and $\partial_{x'} Q(x,x')|_{x'=x}=\sigma'(x)$:
\beq
Q(x,x') = 2\sigma(x) +(x'-x)\sigma'(x)+{4\over 3}\,{(x'-x)^2\over 2}\,S(x)    +O(x'-x)^3
\eeq
where $S(x)$ is called projective connection at $x$.
We can write:
\beq\label{defQB}
Q(x,x') = 2\sigma(x)+(x'-x)\sigma'(x)+(x-x')^2 A(x,x')
\eeq
where $A(x,x')$ is a polynomial in both variables.
We have:
\bea
{B(x,x')\over dx dx'} &=& {1\over 2(x-x')^2}
+ {\sigma(x)\over 2(x-x')^2\ssq{x}\ssq{x'}} \cr
&& + {\sigma'(x)\over 4(x'-x)\ssq{x}\ssq{x'}}
+ {A(x,x')\over 4\ssq{x}\ssq{x'}}
\eea

\newsection{Loop equations}\label{loopeqlargeN}

Now, we will introduce a method for computing the $W^{(h)}_k$'s.
It is based on the so-called loop equations or Schwinger-Dyson equations, i.e. invariance
of the integral \ref{Zdef} under local infinitesimal change of variable.

\subsection{Useful notations}

Let $K=\{2,\dots,k\}$. For any $j\leq k-1$ we denote:
\beq
K_j := \{ I\subset K \,\,\, / \, \# I=j\}
\eeq
and for any subset $I\in K_j$ we define:
\beq
I=\{i_1,i_2,\dots,i_j\} \quad \longrightarrow
\quad
x_{I}:= x_{i_1},x_{i_2},\dots,x_{i_j}
\eeq
as well as:
\beq
\ssq{x_I}:=\prod_{l=1}^j \ssq{x_{i_j}}
\eeq
and
\beq
dx_I:=\prod_{l=1}^j dx_{i_j}
\eeq

\bigskip

\subsection{Loop equations}

The invariance of the matrix integral \ref{Zdef} under the change of variable $M\to M+\eta \delta M$ (see \cite{ZJDFG,eynard, eynardchain}for detailed derivations):
\beq
\delta M = {1\over x_1-M} \prod_{j=2}^{k} \tr {1\over x_j-M}
\eeq
implies, to first order in $\eta$:
\bea\label{chvar}
k=1:&& \overline{W}_1(x_1)^2 + {1\over N^2} \overline{W}_{2}(x_1,x_1) = V'(x_1) \overline{W}_1(x_1) - \overline{U}_1(x_1) \cr
k\geq 2:
&& 2\overline{W}_1(x_1) \overline{W}_k(x_1,\dots,x_k)
+ {1\over N^2} \overline{W}_{k+1}(x_1,x_1,x_2,\dots,x_k) \cr
&&  + \sum_{j=1}^{k-2}
\sum_{I\in K_j}
\overline{W}_{j+1}(x_1,x_{I})
 \overline{W}_{k-j}(x_1,x_{K-I}) \cr
&& + \sum_{j=2}^{k} {\d \over \d x_j} {\overline{W}_{k-1}(x_2,\dots,x_j,\dots,x_k)
 -\overline{W}_{k-1}(x_2,\dots,x_1,\dots,x_k)\over x_j-x_1} \cr
&& = V'(x_1) \overline{W}_k(x_1,\dots,x_k) - \overline{U}_k(x_1;x_2,\dots,x_k)
\eea
i.e., to leading order in $1/N^2$ we have:
\bea
k=1:&& {W}_1(x_1)^2  = V'(x_1) {W}_1(x_1) - {U}_1(x_1) \cr
k\geq 2:
&& 2{W}_1(x_1) {W}_k(x_1,\dots,x_k) \cr
&&  + \sum_{j=1}^{k-2}
\sum_{I\in K_j} {W}_{j+1}(x_1,x_{I}){W}_{k-j}(x_1,x_{K-I}) \cr
&& + \sum_{j=2}^{k} {\d \over \d x_j} {{W}_{k-1}(x_2,\dots,x_j,\dots,x_k)
 -{W}_{k-1}(x_2,\dots,x_1,\dots,x_k)\over x_j-x_1} \cr
&& = V'(x_1) {W}_k(x_1,\dots,x_k) - {U}_k(x_1;x_2,\dots,x_k)
\eea
Notice that it implies \ref{W1} for $k=1$.

Now assume $k\geq 2$, and using \ref{W1}, we rewrite:
\bea\label{loopeqleading}
&& M(x_1) \ssq{x_1} W_k(x_1,\dots,x_k) \cr
&& =
  \sum_{j=1}^{k-2} \sum_{I\in K_j}
W_{j+1}(x_1,x_{I})
 W_{k-j}(x_1,x_{K-I}) \cr
&& + \sum_{j=2}^{k} {\d \over \d x_j} {W_{k-1}(x_2,\dots,x_j,\dots,x_k)
 -W_{k-1}(x_2,\dots,x_1,\dots,x_k)\over x_j-x_1} \cr
&& + U_k(x_1;x_2,\dots,x_k)
\eea

\subsection{Case k=2}

For $k=2$, \ref{loopeqleading} reads:
\bea
M(x_1) \ssq{x_1} W_2(x_1,x_2)
& = & {\d \over \d x_2} \left({W_{1}(x_2) -W_{1}(x_1)\over x_2-x_1}\right)  + U_2(x_1;x_2) \cr
& = &
- {1\over 2}{\d \over \d x_2} \left({M(x_2)\ssq{x_2} -M(x_1)\ssq{x_1}\over x_2-x_1}\right) \cr
&& +{1\over 2}{\d \over \d x_2} \left({V'(x_2) -V'(x_1)\over x_2-x_1}\right) + U_2(x_1;x_2) \cr
& = &
- {1\over 2}{\d \over \d x_2} \left(M(x_1){\ssq{x_2} -\ssq{x_1}\over x_2-x_1}\right) \cr
&& - {1\over 2}{\d \over \d x_2} \left(\ssq{x_2}{M(x_2) -M(x_1)\over x_2-x_1}\right) \cr
&& +{1\over 2}{\d \over \d x_2} \left({V'(x_2) -V'(x_1)\over x_2-x_1}\right) + U_2(x_1;x_2) \cr
\eea
which can be written:
\bea
 \ssq{x_1} W_2(x_1,x_2)
& = &
- {1\over 2}{\d \over \d x_2} \left({\ssq{x_2} -\ssq{x_1}\over x_2-x_1}\right)  + {R_2(x_1;x_2)\over M(x_1)} \cr
\eea
where $R_2(x_1;x_2)$ is a polynomial in $x_1$ of degree at most $d-2$.
From \ref{Wkmcirclevanish}, we know that the LHS has no pole at the zeroes of $M$, thus
\beq
R_2(x_1;x_2)= M(x_1) P_2(x_1;x_2)
\eeq
where $P_2(x_1;x_2)$ is a polynomial in $x_1$, of degree $s-2$. We have:
\bea
 \ssq{x_1} W_2(x_1,x_2)
& = &
- {1\over 2}{\d \over \d x_2} \left({\ssq{x_2} -\ssq{x_1}\over x_2-x_1}\right)  + P_2(x_1;x_2) \cr
\eea
In terms of the function $F_2$ introduced in \ref{defFk} we have:
\beq\label{F2}
{F_2(x_1,x_2)\over 4\ssq{x_1}\ssq{x_2}}
=   {1\over 2\ssq{x_1}}\,{\d \over \d x_2} {\sqrt{\sigma(x_2)}\over (x_1-x_2)}
+ {P_2(x_1;x_2)\over \ssq{x_1}}
\eeq
which proves that $F_2$ is a rational function of $x_1$, and by symmetry, it is also a rational function of $x_2$.

Then using \ref{Wkacyclevanishintro} as well as \ref{decompholomform}
and \ref{defCj}, we find $P_2$:
\beq
{F_2(x_1,x_2)\over 4\ssq{x_1}\ssq{x_2}}
=   {1\over 2}\,{\d \over \d x_2}\,{ \sqrt{\sigma(x_2)}\over \ssq{x_1}}\, \left({1\over (x_1-x_2)}-\sum_l C_l(x_2)L_l(x_1)\right)
\eeq
then using \ref{CjLambda}:
\beq
{F_2(x_1,x_2)\over 4\ssq{x_1}\ssq{x_2}}
=   {1\over 2}\,{\d \over \d x_2}\,{dS(x_1,x_2)\over dx_1}
= {B(x_1,x_2)\over dx_1 dx_2} - {1\over 2}{1\over (x_1-x_2)^2}
\eeq
where we recognize $B$ the Bergmann kernel introduced in \ref{defBergmann}.
Finally, we have the two-loop function in the form:
\beq\label{W2}
\encadremath{
W_2(x_1,x_2) = {B(x_1,x_2)\over dx_1 dx_2} - {1\over (x_1-x_2)^2} = -{B(x_1,\overline{x_2})\over dx_1 dx_2}
}\eeq
or, using \ref{BergmanQ}
\beq\label{F2Q}
F_2(x_1,x_2) = {Q(x_1,x_2)\over (x_1-x_2)^2}
\eeq
The result \ref{W2} or \ref{F2} is well known and can be found in many places in the
literature \cite{ekk}.
We have just presented one derivation for completeness.
Now, let us move to higher loop functions.

\bigskip
Remark: we can write
\beq\label{exprFsigma}
F_2(x,x_1)=2{\sigma(x)\over (x-x_1)^2}-{\sigma'(x)\over x-x_1}+A(x,x_1)
\eeq
where $A(x,x_1)$ is a polynomial in both variables.
It implies:
\beq
W_2(x,x)
=
-{\sigma''(x)\over 8\sigma(x)}
+{\sigma'(x)^2\over 16\sigma(x)^2}
+{A(x,x)\over 4\sigma(x)}
\eeq
which is a rational function of $x$, with double poles at the branch-points.

\subsection{k=3}

Starting from \ref{loopeqleading} for $k=3$, i.e.
\bea\label{loopeqk3}
 M(x_1) \ssq{x_1} W_3(x_1,x_2,x_3)
&=&
2 W_{2}(x_1,x_2) W_2(x_1,x_3) \cr
&& +{\d \over \d x_2} {W_2(x_2,x_3) -W_2(x_1,x_3)\over x_2-x_1} \cr
&& +{\d \over \d x_3} {W_2(x_2,x_3) -W_2(x_1,x_2)\over x_3-x_1} \cr
&& + U_3(x_1;x_2,x_3)
\eea
and using the results for $k=2$, we get:
\bea\label{F3rec}
{F_3(x_1,x_2,x_3)\over \ssq{x_2}\ssq{x_3}}
&=&
 {F_{2}(x_1,x_2) F_2(x_1,x_3)\over  \sigma(x_1) M(x_1) \ssq{x_2}\ssq{x_3}} \cr
&& +{8\over M(x_1)}\,{\d \over \d x_2} {W_2(x_2,x_3)\over x_2-x_1} +{8\over M(x_1)}\,{\d \over \d x_3} {W_2(x_2,x_3)\over x_3-x_1} \cr
&& + {8\, U_3(x_1;x_2,x_3)\over M(x_1)}
\eea
i.e. $F_3(x_1,x_2,x_3)$ is a rational function of $x_1$, and by symmetry, it is a rational function of all its arguments.
Expression \ref{loopeqk3} clearly shows that $F_3$ has no pole when $x_1=x_2$ or $x_1=x_3$.
Moreover, from \ref{Wkmcirclevanish}, we know that it has no pole at the zeroes of $M$.
Thus the only possible poles of $F_3$ are at the branch points and at $\infty$.
Notice that only the first term in the RHS of \ref{F3rec} has poles at the branch points.

Before continuing, let us study the case $k>3$.

\subsection{k larger or equal to 3}

Now assume $k> 3$.
We start from \ref{loopeqleading}:
\bea
  \ssq{x_1} W_k(x_1,x_K)
&=&
  \sum_{j=1}^{k-2} \sum_{I\in K_j}
{W_{j+1}(x_1,x_{I}) W_{k-j}(x_1,x_{K-I}) \over M(x_1)}\cr
&& + \sum_{i=2}^{k} {\d \over \d x_i} {W_{k-1}(x_K)
 -W_{k-1}(x_1,x_{K-\{i\} })\over (x_i-x_1)M(x_1)} \cr
&& + {U_k(x_1;x_K)\over M(x_1)}
\eea
and we consider separately the terms corresponding to $j=1$ and $j=k-2$ in the RHS,
and we write $\sum_{I\in K_1}=\sum_{i=2}^k$, we get:

\bea\label{calcul1}
  \ssq{x_1} W_k(x_1,x_K)
&=&
  \sum_{j=2}^{k-3} \sum_{I\in K_j}
{W_{j+1}(x_1,x_{I}) W_{k-j}(x_1,x_{K-I}) \over M(x_1)}\cr
&&  +2\sum_{i=2}^k {\left(W_{2}(x_1,x_i) +{1\over 2 (x_1-x_i)^2}\right) W_{k-1}(x_1,x_{K-I}) \over M(x_1)}\cr
&& + \sum_{i=2}^{k} {\d \over \d x_i} {W_{k-1}(x_K)\over (x_i-x_1)M(x_1)}  + {U_k(x_1;x_K)\over M(x_1)}
\eea
This clearly proves, by induction on $k$, that for all $k\geq 3$, $F_k(x_1,x_K)$ is a rational function of $x_1$ (and by symmetry, of all its arguments),
with poles only at the branch points and at $\infty$.
We have just re-derived it in a way different from \cite{akemanambjorn, akeman}.

\medskip

Now, assume  $k\geq 3$.
Consider the Euclidean division of the polynomial $U_k(x_1;x_2,\dots,x_k)$ by $M(x_1)$:
\beq
U_k(x_1;x_K)= {2^{-k}\over \ssq{x_K}}\,P_k(x_1;x_K)\, M(x_1) + Q_k(x_1;x_K)
\eeq
where $\deg P_k=s-2$ and $\deg Q_k< d-s$.

Thus, we have found that for any $k\geq 3$, we have:
\beq\label{calcul}
  {F_k(x_1,x_K)- P_k(x_1;x_K)\over  \ssq{x_K}}
=
{1\over 2}\,  \sum_{j=1}^{k-2} \sum_{I\in K_j}
{F_{j+1}(x_1,x_{I}) F_{k-j}(x_1,x_{K-I}) \over  \sigma(x_1) M(x_1) \ssq{x_K}} + R_k(x_1;x_K)
\eeq
where $R_k(x_1;x_K)$ is a rational fraction of $x_1$ with no poles at the branch points neither at $\infty$ (it has poles at the zeroes of $M$ and at the $x_i$'s).

\subsection{Cauchy formula}

Cauchy formula gives:
\beq
  {F_k(x_1,x_K)- P_k(x_1;x_K)\over  \ssq{x_K}}
= \Res_{x\to x_1} {dx\over x-x_1} {F_k(x,x_K)- P_k(x;x_K)\over  \ssq{x_K}}
\eeq
where the integrand has poles only at the branch points.
Therefore we may deform the integration contour used to compute the residue, into residues at the branch points only:
\beq
 F_k(x_1,x_K) - P_k(x_1;x_K)
= \sum_{l=1}^{2s} \Res_{x\to a_l} {dx\over x_1-x} \left(F_k(x,x_K)- P_k(x;x_K) \right)
\eeq
using  \ref{calcul} we get the recursion formula for the $F_k$'s, for all $k\geq 3$:
\beq\label{recFkPk}
 F_k(x_1,x_K) - P_k(x_1;x_K)
= {1\over 2}\, \sum_{l=1}^{2s} \Res_{x\to a_l}  \sum_{j=1}^{k-2} \sum_{I\in K_j}
{F_{j+1}(x,x_{I}) F_{k-j}(x,x_{K-I}) \over  \sigma(x) M(x) (x_1-x)}\,dx
\eeq
$P_k(x_1,x_K)$ which is a polynomial in $x_1$ of degree at most $s-2$, is computed with formula \ref{decompholomform}.

Starting from  \ref{Wkacyclevanishintro}, i.e.:
\beq
\oint_{[a_{2l-1},a_{2l}]} {F_k(x_1,x_K)\over \ssq{x_1}}\,dx_1=0
\eeq
we have:
\bea
&& -\oint_{[a_{2l-1},a_{2l}]} {P_k(x_1;x_K)dx_1\over \ssq{x_1}} \cr
&=& \sum_{i=1}^{2s}  \sum_{j=1}^{k-2} \sum_{I\in K_j}
\oint_{[a_{2l-1},a_{2l}]} {dx_1\over \ssq{x_1}} \Res_{x\to a_i}
{F_{j+1}(x,x_{I}) F_{k-j}(x,x_{K-I}) \over 2(x_1-x)M(x)\sigma(x)}dx\cr
&=& {1\over 2i\pi}\,\sum_{i=1}^{s}  \sum_{j=1}^{k-2} \sum_{I\in K_j}
\oint_{[a_{2l-1},a_{2l}]} {dx_1\over \ssq{x_1}} \oint_{[a_{2i-1},a_{2i}]}
{F_{j+1}(x,x_{I}) F_{k-j}(x,x_{K-I}) \over 2(x_1-x)M(x)\sigma(x)}dx\cr
\eea
Notice that for $i=l$, the contour of integration of $x_1$ encloses the contour of $x$.
We may exchange the position of the two contours, by picking a residue at $x_1=x$, thus:
\bea
&& -\oint_{[a_{2l-1},a_{2l}]} {P_k(x_1;x_K)dx_1\over \ssq{x_1}} \cr
&=& {1\over 2i\pi}\,\sum_{i=1}^{s}  \sum_{j=1}^{k-2} \sum_{I\in K_j}
\oint_{[a_{2i-1},a_{2i}]} \!\!\!\!\!\!\!\!\!\!{F_{j+1}(x,x_{I}) F_{k-j}(x,x_{K-I}) \over 2 M(x)\sigma(x)}\,dx \oint_{[a_{2l-1},a_{2l}]} {dx_1\over (x_1-x)\ssq{x_1}} \cr
&& + \sum_{j=1}^{k-2} \sum_{I\in K_j}
\oint_{[a_{2l-1},a_{2l}]}
{F_{j+1}(x,x_{I}) F_{k-j}(x,x_{K-I}) \over 2\, M(x)\ssq{x}\,\sigma(x)}dx\cr
\eea
Using the function $C_l(x)$ introduced in \ref{defCj}, we have:
\bea
&& -\oint_{[a_{2l-1},a_{2l}]} {P_k(x_1;x_K)dx_1\over \ssq{x_1}} \cr
&=& \sum_{i=1}^{s}  \sum_{j=1}^{k-2} \sum_{I\in K_j}
\oint_{[a_{2i-1},a_{2i}]} {F_{j+1}(x,x_{I}) F_{k-j}(x,x_{K-I}) \over 2M(x)\sigma(x)}\,dx
\left( C_l(x)+{\delta_{i,l}\over \ssq{x}}\right) \cr
\eea
and thus we have computed $P_k$:
\bea
 - P_k(x_1;x_K)
&=& {1\over 2i\pi}\,\sum_{i=1}^{s}  \sum_{j=1}^{k-2} \sum_{I\in K_j}
\oint_{[a_{2i-1},a_{2i}]} {F_{j+1}(x,x_{I}) F_{k-j}(x,x_{K-I}) \over 2M(x)\sigma(x)}\,dx \cr
&& \qquad\qquad \left( {L_i(x_1)\over \ssq{x}}+\sum_{l=1}^{s-1} C_l(x)L_l(x_1)\right) \cr
\eea

\subsection{The recursion relation}

That gives the recursion relation for the $F_k$'s:
\beq\label{mainrecFk}
\encadremath{
\begin{array}{l}
\displaystyle F_k(x_1;x_K)
= {1\over 2i\pi}\,\sum_{i=1}^{s}  \sum_{j=1}^{k-2} \sum_{I\in K_j} \cr
\displaystyle \oint_{[a_{2i-1},a_{2i}]} {F_{j+1}(x,x_{I}) F_{k-j}(x,x_{K-I}) \over 2M(x)\sigma(x)}\,dx
\left( {1\over x_1-x}-{L_i(x_1)\over \ssq{x}}-\sum_{l=1}^{s-1} C_l(x)L_l(x_1)\right) \cr
\end{array}
}\eeq
where it is important to remember that the term inside the bracket is analytical when $x$ approaches $a_{2i-1}$ or $a_{2i}$.
This allows to write the contour integrals as the sum of two residues around $a_{2i-1}$ and $a_{2i}$.

\bigskip

It is interesting and more intrinsic to rewrite \ref{mainrecFk} in terms of multi-linear differentials $G_k$ on the hyperelliptical curve.

First, notice that a contour around $a_i$ in the hyperelliptical curve ($\tau_i(x)=\sqrt{x-a_i}$ around $0$)
is twice the contour in the complex plane ($x$ around $a_i$), i.e. we will have an extra factor $2$ in the denominator.

Then, notice that $G_2$ and $F_2$ differ by a term, which is an even function of the local parameter $\tau_i(x)$, i.e. which does not contribute to residues near the branch-points
(this can be checked separately for $k=3$ and $k>3$).

Thus we get the recursion relation for the $G_k$'s:
\beq\label{mainrecGk}
\encadremath{
 G_k(x_1;x_K)
= \sum_{i=1}^{2s}  \sum_{j=1}^{k-2} \sum_{I\in K_j}  \Res_{a_{i}\in\Sigma} {G_{j+1}(x,x_{I}) G_{k-j}(x,x_{K-I}) \over 2y(x) dx }\, dS_{i}(x_1,x)
}\eeq
where now the residues are computed on the hyperelliptical surface (i.e. extra factor $2$ in the denominator),
and $dS_i(x,x')$ is the abelian differential of the third kind introduced in \ref{defdSi}.

\bigskip

That recursion relation allows to compute $W_k$ in a tree-like recursion from residues of lower loop-functions.

\subsection{Solution of the recursion relation as cubic-Feynmann trees}

Equation \ref{mainrecGk} is conveniently represented with diagrams ''a la Feynmann''.

Let us represent the $k$-loop correlation function $G_k(x_1,\dots,x_k)$ as a black disk with $k$ legs:
\beq
\begin{array}{r}
{\epsfxsize 3cm\epsffile{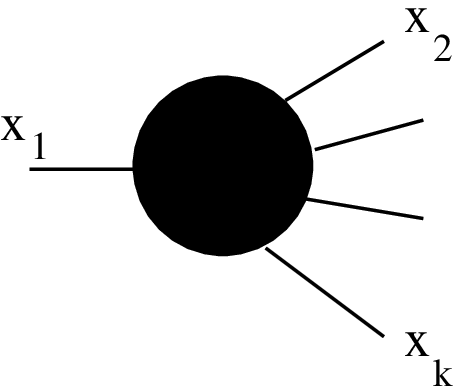}}
\end{array}
 :=G_k(x_1,\dots,x_k)
\eeq
\medskip
and introduce the following Feynmann rules:

\medskip

\begin{center}
\begin{tabular}{|r|l|}\hline
Arrowed propagator:&
$
\begin{array}{r}
{\epsfxsize 2cm\epsffile{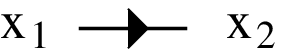}}
\end{array}
:=dS_i(x,x')
$
\cr\hline
Vertex:&
$
\begin{array}{r}
{\epsfxsize 2cm\epsffile{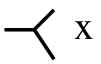}}
\end{array}
 :=1/(2\, y(x)\, dx)
$
\cr\hline
Non-arrowed propagator&
$
\begin{array}{r}
{\epsfxsize 2cm\epsffile{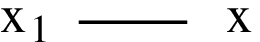}}
\end{array}
:=
\begin{array}{r}
{\epsfxsize 2cm\epsffile{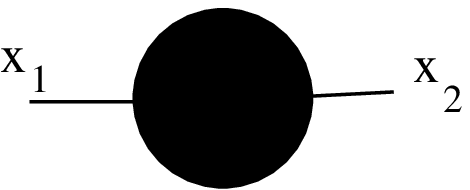}}
\end{array}
=G_2(x_1,x_2)
$
\cr
=$2$-loop correlator:&
\cr
\hline
\end{tabular}
\end{center}

\medskip

Then \ref{mainrecGk} can be represented as follows
$$
{\epsfysize 4cm\epsffile{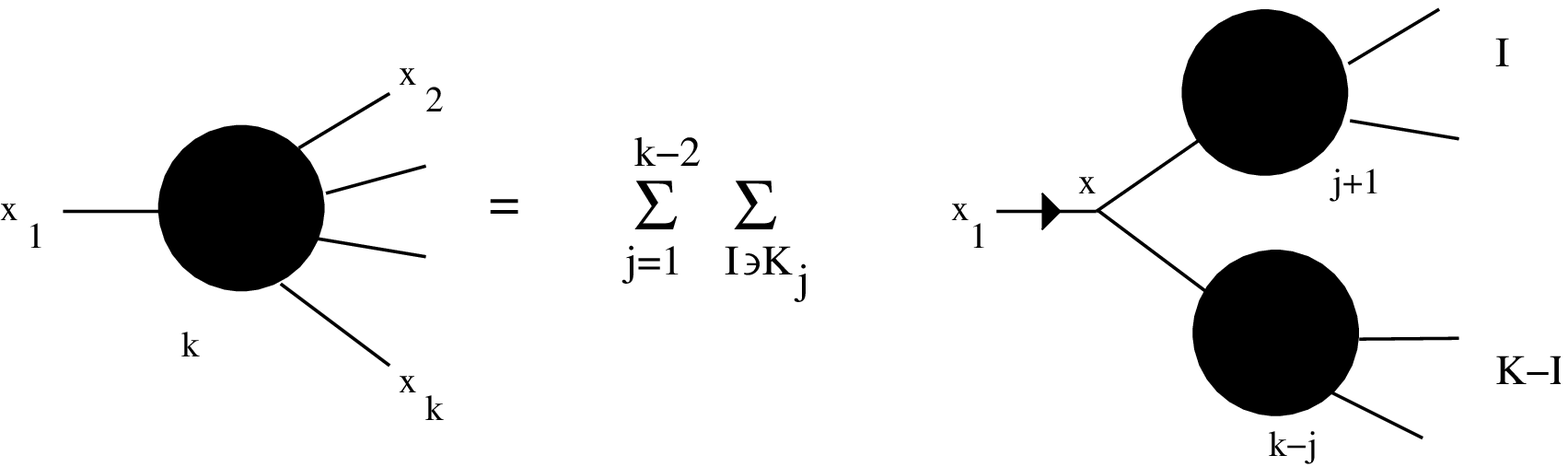}}
$$

Whose solution is clearly that the $k$-loop correlator $G_k$ is the sum over all plane binary trees with $1$ root and $k-1$ leaves, with a skeleton made of oriented arrows (oriented from root toward leaves), and whose $k-1$ leaves are non-arrowed propagators finishing at the $x_j$'s with $2\leq j\leq k$.

Notice that two trees which differ only by the ordering of branches at a vertex give the same contribution to $G_k$, so that instead of summing over plane trees,
one can sum over non-plane trees, with a factor $2^{k-2}$.

Let ${\cal T}_k$ be the set of plane rooted binary trees with $k-1$ labeled leaves ($x_2,\dots,x_k$).
and let $\overline{{\cal T}}_k$ be the set of non-plane rooted binary trees with $k-1$ labeled leaves ($x_2,\dots,x_k$).
We have:
\beq\label{nbtrees}
N_{k+2}:={\rm Card}\, {\cal T}_{k+2} = k+1!\, C_k  = {2k!\over k!} = 2^k\,\,(2k-1)!!
\eeq
where $C_k$ is the Catalan number which enumerates plane trees. And:
\beq\label{nbtreesbar}
\overline{N}_{k+2}:={\rm Card}\, \overline{{\cal T}}_{k+2}=2^{-k}{\rm Card}\, {\cal T}_{k+2} = 2^{-k}\,{2k!\over k!} = (2k-1)!!
\eeq

For any given tree $T\in {\cal T}_k$, with root $x_1$, leaves $x_j$ ($j=2,\dots, k$), and with $k-2$ vertices noted $x'_v$ ($v=1,\dots,k-2$), so that its inner edges are of the form
$v_1\to v_2$ and its outer edges are of the form $v\to j$, we define the weight of $T$ as:
\bea\label{defweightW}
{\cal W}(T)
&:=&
\prod_{{\rm vertex}\, v=1}^{k-2} \sum_{i_v=1}^{2s} \Res_{x'_v\to a_{i_v}} {1\over 2y(x'_v)dx'_v} \cr
&& \prod_{{\rm inner\, edges}\, v\to w} dS_{i_v}(x'_v,x'_w)
\prod_{{\rm outer\, edges}\, v\to j} G_2(x'_v,x_j) \cr
\eea

Thus we have:
\beq\label{GkcalW}
\encadremath{
G_k(x_1,\dots,x_k)
 =  2^{k-2} \sum_{T\in \overline{{\cal T}}_k}\quad {\cal W}(T)
 =   \sum_{T\in {\cal T}_k}\quad {\cal W}(T)
}\eeq

$G_3$ is thus given by $\overline{N}_3=1$ tree, $G_4$ is the sum of $\overline{N}_4=3$ diagrams, $G_5$ is the sum of $\overline{N}_5=15$ diagrams,...

\subsection{Example: 3-point function}

As an example, let us carry out explicitly the computation for the 3-point function.

Diagrammatically, we have:
$$
{\epsfysize 4cm\epsffile{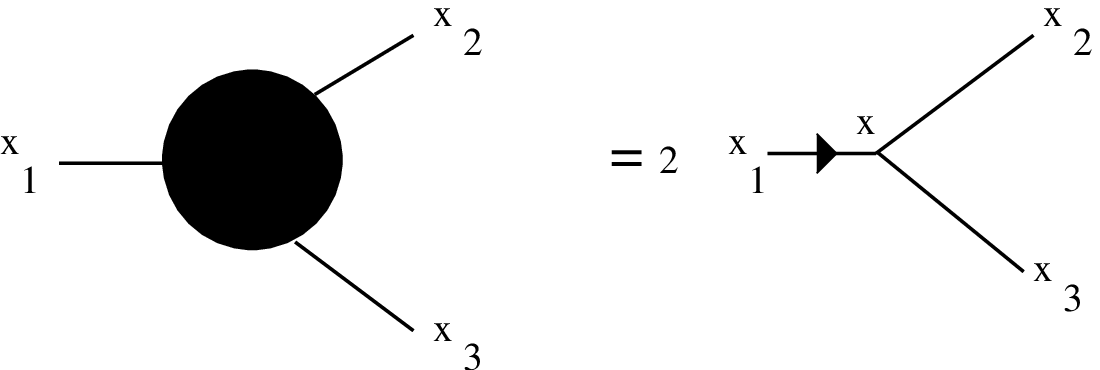}}
$$

Eq.\ref{recFkPk} can be written for $k=3$:
\beq\label{recF3}
\begin{array}{rcl}
F_3(x_1,x_2,x_3)
& = & \displaystyle\sum_{i=1}^{2s} \mathop{\rm Res}_{a_i}
dx\,{F_{2}(x,x_2) F_{2}(x,x_{3})\over M(x)\sigma(x)\sigma'(x)}{\sigma'(x)\over (x-x_1) }
+ P_3(x_1,x_2,x_3)  \cr
\end{array}
\eeq
where $P_3$ is a polynomial in $x_1$.
Using \ref{exprFsigma}, notice that $F_2(a_i,x)={\sigma'(a_i)\over x-a_i}+A(a_i,x)$ is finite, and that:
\beq
{\sigma'(x)\over x-x_1}
 = 2{\sigma(x)\over (x-x_1)^2}-F_2(x,x_1)+A(x,x_1)
\eeq
i.e.
\beq
\begin{array}{rcl}
F_3(x_1,x_2,x_3)
& = &  \displaystyle 2\sum_{i=1}^{2s} \mathop{\rm Res}_{a_i}
dx\,{F_{2}(x,x_2) F_{2}(x,x_{3})\over M(x)(x-x_1)^2\sigma'(x)} \cr
&  & - \displaystyle\sum_{i=1}^{2s} \mathop{\rm Res}_{a_i}
dx\,{F_{2}(x,x_1) F_{2}(x,x_2) F_{2}(x,x_{3})\over M(x)\sigma(x)\sigma'(x)} \cr
&  &  +\displaystyle\sum_{i=1}^{2s} \mathop{\rm Res}_{a_i}
dx\,{A(x,x_1) F_{2}(x,x_2) F_{2}(x,x_{3})\over M(x)\sigma(x)\sigma'(x)}
+ R_3(x_1,x_2,x_3)  \cr
\end{array}
\eeq
The first line has no residue at the branch-points (indeed, using \ref{exprFsigma}, notice that $F_2(a_i,x)$ is finite),
and the last line is a polynomial in $x_1$ (indeed $A(x,x_1)$ is), therefore:
\beq
\begin{array}{rcl}
F_3(x_1,x_2,x_3)
& = &  \displaystyle -\sum_{i=1}^{2s} \mathop{\rm Res}_{a_i}
dx\,{F_{2}(x,x_1) F_{2}(x,x_2) F_{2}(x,x_{3})\over M(x)\sigma(x)\sigma'(x)}
+ \td{R}_3(x_1,x_2,x_3)  \cr
& = &  \displaystyle -\sum_{i=1}^{2s}
{F_{2}(a_i,x_1) F_{2}(a_i,x_2) F_{2}(a_i,x_{3})\over M(a_i)\sigma'(a_i)^2}
+ \td{R}_3(x_1,x_2,x_3)  \cr
\end{array}
\eeq
where $\td{R}_3(x_1,x_2,x_3)$ is a polynomial in $x_1$, of degree at most $s-2$.
Condition \ref{Wkacyclevanishintro} implies that $\td{R}_3(x_1,x_2,x_3)=0$, thus:
\beq\label{F3result}
\encadremath{
\begin{array}{rcl}
F_3(x_1,x_2,x_3)
& = &  \displaystyle - \sum_{i=1}^{2s}
{F_{2}(a_i,x_1) F_{2}(a_i,x_2) F_{2}(a_i,x_{3})\over M(a_i)\sigma'(a_i)^2}   \cr
\end{array}
}\eeq
This is a generalization of what was found in \cite{akemanambjorn, akeman, ACKM} for the one-cut case $s=1$.

\bigskip
Let us redo this computation in a more intrinsic way.
Start from \ref{mainrecGk} for $k=3$:
\bea
G_3(x_1,x_2,x_3)
&=& 2\sum_{i=1}^{2s}   \Res_{a_{i}} {G_{2}(x,x_2) G_{2}(x,x_3) \over 2y(x) dx }\, dS_i(x_1,x) \cr
\eea
Notice that both $dS_i(x_1,x)$ and $y(x)$ have a simple zero at $x=a_i$ thus,
\bea
{dS_i(x_1,x)\over y(x)}
&=&{d_x dS_i(x_1,x)\over dy(x)} + O(\sqrt{x-a_i}) \cr
&=& {B(x_1,x)+B(x_1,\ovl{x})\over dy(x)} + O(\sqrt{x-a_i}) \cr
&=& 2{G_2(x_1,x)\over dy(x)} + O(\sqrt{x-a_i})
\eea

This implies:
\beq
\encadremath{
\begin{array}{rcl}
G_3(x_1,x_2,x_3)
& = & 2\displaystyle\sum_{i=1}^{2s}\mathop{\rm Res}_{a_i}
{G_2(x,{x_2}) G_2(x,{x_3}) G_2(x,x_1)  \over dx\, dy  } \cr
\end{array}
}\eeq
This agrees with \cite{Kric, Marco} (our $dy$ is half the $dy$ of \cite{Marco}).
One can also write:
\beq
\begin{array}{rcl}
W_3(x_1,x_2,x_3)
& = & 2 \displaystyle\sum_{i=1}^{2s}\mathop{\rm Res}_{a_i}
W_2(x,{x_2}) W_2(x,{x_3})  W_2(x,x_1)  {dx^2\over dy } \cr
\end{array}
\eeq

\subsection{Example: 4 point function}

Diagrammatically, we have:
$$
{\epsfxsize 14cm\epsffile{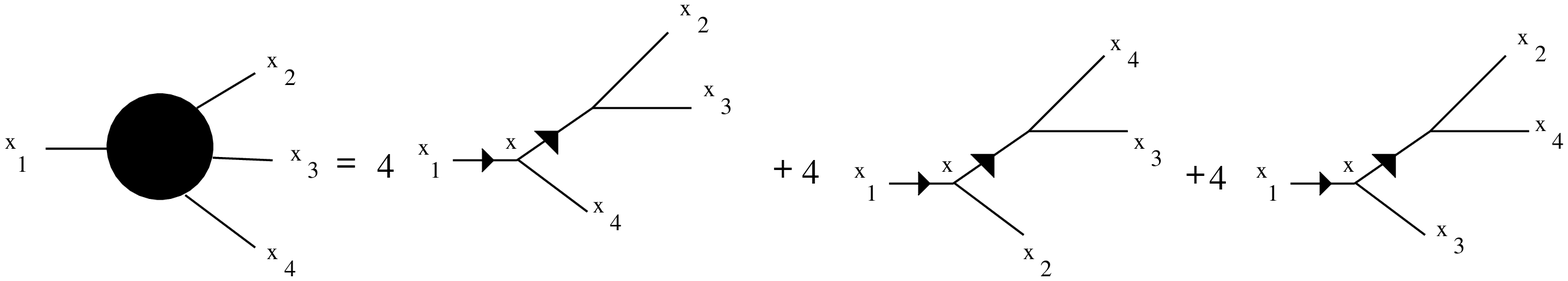}}
$$

Explicit computation of \ref{mainrecFk} for $k=4$ gives:
\beq
\begin{array}{rcl}
F_4(x_1,x_2,x_3,x_4)
&=& -\displaystyle\sum_{i\neq j=1}^{2s} {F_2(x_3,a_j)F_2(x_4,a_j)\over M(a_j)\sigma'(a_j)^2}
F_2(a_i,a_j){F_{2}(x_2,a_i)F_{2}(x_1,a_i)\over M(a_i)\sigma'(a_i)^2}  \cr
&& -\displaystyle\sum_{i\neq j=1}^{2s} {F_2(x_3,a_j)F_2(x_2,a_j)\over M(a_j)\sigma'(a_j)^2}
F_2(a_i,a_j){F_{2}(x_4,a_i)F_{2}(x_1,a_i)\over M(a_i)\sigma'(a_i)^2}  \cr
&& -\displaystyle\sum_{i\neq j=1}^{2s} {F_2(x_2,a_j)F_2(x_4,a_j)\over M(a_j)\sigma'(a_j)^2}
F_2(a_i,a_j){F_{2}(x_3,a_i)F_{2}(x_1,a_i)\over M(a_i)\sigma'(a_i)^2}  \cr
&& -\displaystyle\sum_{i=1}^{2s} {(F_2(x_3,a_i)F_2(x_4,a_i)F_2(a_i,x_2)F_2(a_i,x_1))'\over M(a_i)^2\sigma'(a_i)^3} \cr
&& +3\displaystyle\sum_{i=1}^{2s} {F_2(x_3,a_i)F_2(x_4,a_i)F_2(a_i,x_2)F_2(a_i,x_1)\over M(a_i)^2\sigma'(a_i)^3} \cr
&& \qquad\qquad\displaystyle\left( {M'(a_i)\over M(a_i)} -{A(a_i,a_i)\over \sigma'(a_i)}+{5\sigma''(a_i)\over 6\sigma'(a_i)} \right)  \cr
\end{array}
\eeq

\newsection{Higher genus}\label{loopeqhighergenus}

Now, we don't drop the $1/N^2$ term in \ref{chvar}, and we expand to order $h$:
\bea\label{loopeqgenush}
&& 2\sum_{m=0}^h W_1^{(h-m)}(x_1) W^{(m)}_k(x_1,\dots,x_k) \cr
&& +  W^{(h-1)}_{k+1}(x_1,x_1,x_2,\dots,x_k) \cr
&&  + \sum_{m=0}^{h} \sum_{j=1}^{k-2}
\sum_{I\in K_j}
W^{(m)}_{j+1}(x_1,x_{I})
 W^{(h-m)}_{k-j}(x_1,x_{K-I}) \cr
&& + \sum_{j=2}^{k} {\d \over \d x_j}{ W^{(h)}_{k-1}(x_2,\dots,x_j,\dots,x_k)
 -W^{(h)}_{k-1}(x_2,\dots,x_1,\dots,x_k)\over x_j-x_1} \cr
&& = V'(x_1) W^{(h)}_k(x_1,\dots,x_k) - U^{(h)}_k(x_1;x_2,\dots,x_k)
\eea
thus:
\bea\label{loopeqbisgenush}
&& M(x_1)\ssq{x_1}\, W^{(h)}_k(x_1,x_K) \cr
&& = 2\sum_{m=0}^{h-1} W_1^{(h-m)}(x_1) W^{(m)}_k(x_1,x_K)  +  W^{(h-1)}_{k+1}(x_1,x_1,x_K) \cr
&&  + \sum_{m=0}^{h} \sum_{j=1}^{k-2}
\sum_{I\in K_j}
W^{(m)}_{j+1}(x_1,x_{I})
 W^{(h-m)}_{k-j}(x_1,x_{K-I}) \cr
&& + \sum_{j=2}^{k} {\d \over \d x_j}{ W^{(h)}_{k-1}(x_K)
 -W^{(h)}_{k-1}(x_1,x_{K-\{j\}})\over x_j-x_1}
 + U^{(h)}_k(x_1;x_K) \cr
\eea

In particular for $k=1$, \ref{loopeqbisgenush} reads :
\beq
M(x_1)\ssq{x_1}\, W^{(h)}_1(x_1)
 = \sum_{m=1}^{h-1} W_1^{(h-m)}(x_1) W^{(m)}_1(x_1)  +  W^{(h-1)}_{2}(x_1,x_1)  + U^{(h)}_1(x_1)
\eeq

It is easy to prove, by double recursion on $k$ and $h$, that
\beq
F_k^{(h)}(x_1,\dots,x_k) = 2^k\,W_k^{(h)}(x_1,\dots,x_k)\,\prod_{i=1}^k \ssq{x_i}
\eeq
is a rational function.
Introduce the Euclidean division of the polynomial $U^{(h)}_k(x_1;x_K)$ by $M(x_1)$:
\beq
U^{(h)}_k(x_1;x_K)= {2^{-k}\over \ssq{x_K}}\, P^{(h)}_k(x_1;x_K)\, M(x_1) + Q_k^{(h)}(x_1;x_K)
\eeq
where $\deg P_k^{(h)}=s-2$ and $\deg Q_k^{(h)}< d-s$.
\ref{loopeqbisgenush} becomes:
\bea\label{loopeqtergenush}
&& \ssq{x_1}\, W^{(h)}_k(x_1,x_K) - {2^{-k}\over \ssq{x_K}}\, P^{(h)}_k(x_1;x_K) \cr
&& = 2\sum_{m=0}^{h-1} {W_1^{(h-m)}(x_1) W^{(m)}_k(x_1,x_K)\over M(x_1)}  +  {W^{(h-1)}_{k+1}(x_1,x_1,x_K)\over M(x_1)} \cr
&&  + \sum_{m=0}^{h} \sum_{j=1}^{k-2}
\sum_{I\in K_j}
{W^{(m)}_{j+1}(x_1,x_{I}) W^{(h-m)}_{k-j}(x_1,x_{K-I})\over M(x_1)} \cr
&& + \sum_{j=2}^{k} {\d \over \d x_j}{ W^{(h)}_{k-1}(x_K)
 -W^{(h)}_{k-1}(x_1,x_{K-\{j\}})\over (x_j-x_1)\,M(x_1)}
 + Q^{(h)}_k(x_1;x_K) \cr
\eea
where the LHS is a rational function of $x_1$ with poles only at the branch points, therefore:
\bea
&& \ssq{x_1}\, W^{(h)}_k(x_1,x_K) - {2^{-k}\over \ssq{x_K}}\, P^{(h)}_k(x_1;x_K) \cr
&&= \Res_{x\to x_1} {dx\over x-x_1}\, \left( \ssq{x}\, W^{(h)}_k(x,x_K) - P^{(h)}_k(x;x_K)\right) \cr
&&= \sum_{i=1}^{2s}\,\Res_{x\to a_i} {dx\over x_1-x}\, \left( \ssq{x}\, W^{(h)}_k(x,x_K) - P^{(h)}_k(x;x_K)\right) \cr
&&= \sum_{i=1}^{2s}\,\Res_{x\to a_i} {dx\over x_1-x}\, \left(
2\sum_{m=0}^{h-1} {W_1^{(h-m)}(x) W^{(m)}_k(x,x_K)\over M(x)}  +  {W^{(h-1)}_{k+1}(x,x,x_K)\over M(x)} \right) \cr
&&+ \sum_{i=1}^{2s}\,\Res_{x\to a_i} {dx\over x_1-x}\, \left(
 \sum_{m=0}^{h} \sum_{j=1}^{k-2}\sum_{I\in K_j}
{W^{(m)}_{j+1}(x,x_{I}) W^{(h-m)}_{k-j}(x,x_{K-I})\over M(x)} \right) \cr
&&+ \sum_{i=1}^{2s}\,\Res_{x\to a_i} {dx\over x_1-x}\, \left(
 \sum_{j=2}^{k} {\d \over \d x_j}{ W^{(h)}_{k-1}(x_K)
 -W^{(h)}_{k-1}(x,x_{K-\{j\}})\over (x_j-x)\,M(x)}
 + {Q^{(h)}_k(x;x_K)\over M(x)}\right) \cr
\eea
Two terms in the last line have no pole at $x_1=a_i$,
and the other term in the last line combines with other lines so as to transform $W_2$ in $F_2$.
Thus we get:
\bea
 F^{(h)}_k(x_1,x_K)
&=& {1\over 2}\,\sum_{i=1}^{2s}\,\Res_{x\to a_i} {dx\over x_1-x}\,
 \sum_{m=0}^{h} \sum_{j=0}^{k-1} (1-\delta_{m,0}\delta_{j,0}-\delta_{m,h}\delta_{j,k-1}) \cr
 && \,\sum_{I\in K_j}\,{F^{(m)}_{j+1}(x,x_{I}) F^{(h-m)}_{k-j}(x,x_{K-I})\over M(x)\sigma(x)}  \cr
&&+ {1\over 2}\,\sum_{i=1}^{2s}\,\Res_{x\to a_i} {dx\over x_1-x}\,  {F^{(h-1)}_{k+1}(x,x,x_K)\over M(x)\sigma(x)}  \cr
&& +P^{(h)}_k(x_1;x_K)\cr
\eea
where $P^{(h)}_k(x_1;x_K)$ is obtained from \ref{decompholomform} in a way very similar to what we have done to leading order.
Finally we find:
\beq\label{mainrecGkh}
\encadremath{
\begin{array}{lll}
 G^{(h)}_k(x_1,x_K)
&=&
\sum_{i=1}^{2s}\,\Res_{x\to a_i} dS_i(x_1,x)\,  {G^{(h-1)}_{k+1}(x,x,x_K)\over y(x)dx}  \cr
&& + 2\sum_{i=1}^{2s}\,\Res_{x\to a_i} \sum_{m=0}^{h-1}dS_i(x_1,x)\,
{G_1^{(h-m)}(x) G^{(m)}_k(x,x_K)\over y(x)dx} \cr
&&+ \sum_{i=1}^{2s}\,\Res_{x\to a_i} \sum_{m=0}^{h} \sum_{j=1}^{k-2}\sum_{I\in K_j} \cr
&& \qquad \quad dS_i(x_1,x)\,{G^{(m)}_{j+1}(x,x_{I}) G^{(h-m)}_{k-j}(x,x_{K-I})\over y(x)dx}  \cr
\end{array}
}\eeq
where one should notice that the first line correspond to $j=0$ and $j=k-1$ in the second line.

Let us represent the order $N^{-2h}\,\,$ $k$-loop correlation function $G^{(h)}_k(x_1,\dots,x_k)$ as a black disk with $k$ legs and $h$ holes:
\beq
\begin{array}{r}
{\epsfxsize 3cm\epsffile{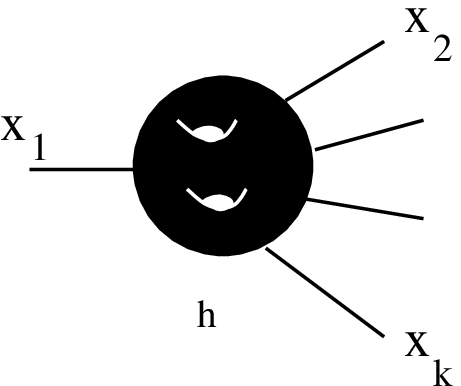}}
\end{array}
 :=G^{(h)}_k(x_1,\dots,x_k)
\eeq
\medskip

Using the Feynmann rules introduced above, \ref{mainrecGkh} can be represented as:
$$
{\epsfxsize 15cm\epsffile{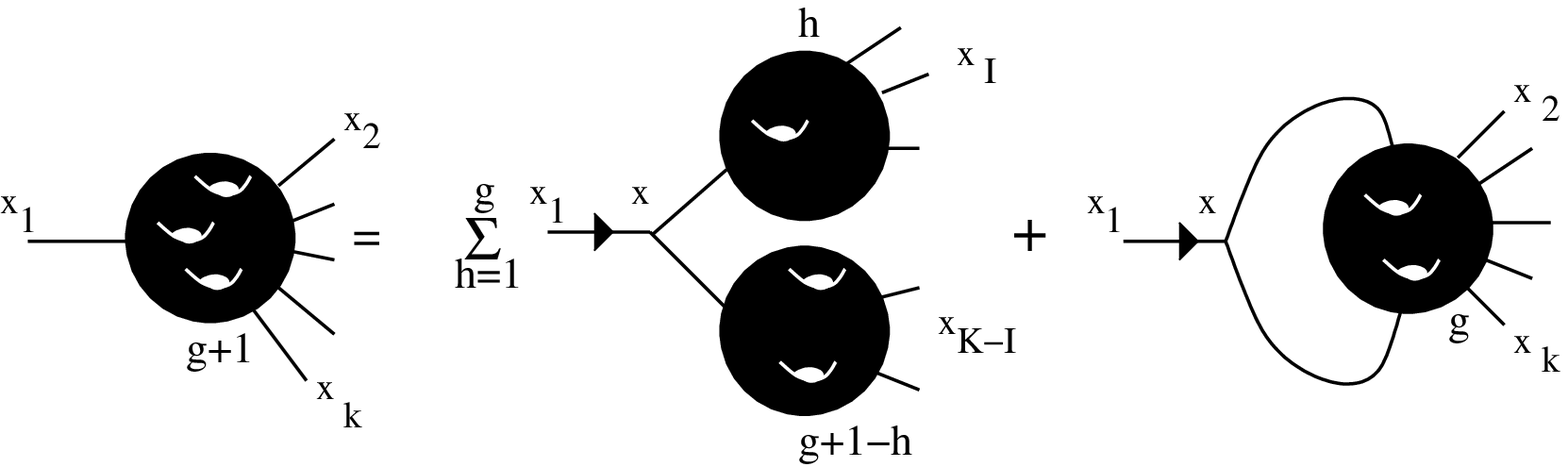}}
$$
which means that $G_k^{(h)}$ is obtained by summing over all Feynmann graphs with $k$ external legs and $h$ loops.

\begin{figure}
$$
{\epsfxsize 14cm\epsffile{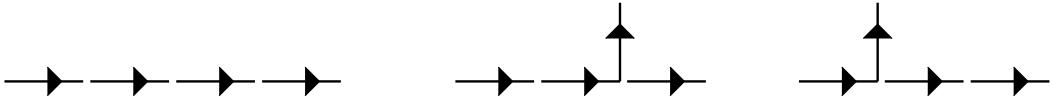}}
$$
\caption{The 3 skeleton trees contributing to ${\cal T}_{2}^{(2)}$, i.e. with $k+2h-2=4$ edges.}\label{figT22trees}
\end{figure}

\begin{figure}
$$
{\epsfxsize 13cm\epsffile{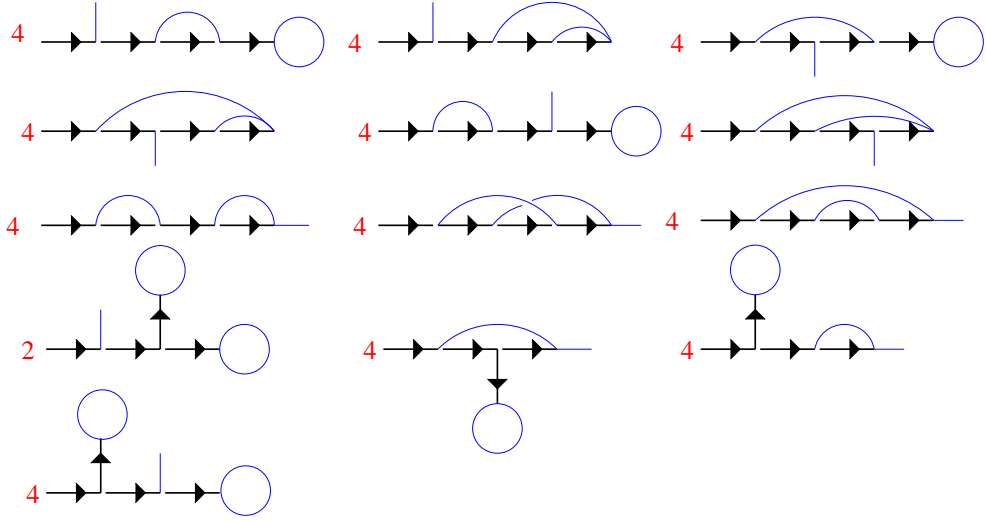}}
$$
\caption{All the possible ways of drawing $k-1=1$ external leg and $h=2$ inner edges,
so that the graphs are trivalent, and that inner edges never connect different branches.
Notice that all but one graph have symmetry factor $4$, and one has $2$.}\label{figT22}
\end{figure}

The set ${\cal T}_k^{(h)}$ of all possible graphs with $k$ external legs, and with $h$ loops, can be described as follows:
First, draw all rooted skeleton trees (i.e. trees whose vertices have valence 1,2 or 3), containing $k+2h-2$ edges.
Draw arrows on the edges, oriented from root toward leaves (see fig \ref{figT22trees}).
Then draw, in all possible ways, $k-1$ external legs, and $h$ inner edges, with the constraint that all the vertices of the whole graph have valence $3$,
and so that an inner edge can be drawn only between a vertex and one of its descendents (inner edges can never connect different branches of the tree),
 see fig \ref{figT22} for the example $k=2$, $h=2$.
Then, each such graph has a symmetry factor.

We have (see appendix \ref{appNkh}):
\beq
N_k^{(h)}:={\rm Card}\, {\cal T}_k^{(h)} = s_h\, (k-1)!\,4^{k-1}\,\pmatrix{{3(h-1)\over 2}+k-1\cr k-1}
\eeq
where $s_h=N_1^{(h)}$ is the number of one-leg graphs in a usual $\phi^3$ field theory.
The generating function $s(x)=\sum_{h=1}^\infty s_h x^{h-1}$ is computed in appendix \ref{appNkh} in terms of Airy function.
We have:
\beq
s_1=1\virg s_2=5\virg s_3=60\virg \dots
\eeq
In particular for genus $h=1$:
 \beq
N_k^{(1)}:={\rm Card}\, {\cal T}_k^{(1)} = 4^{k-1}\,(k-1)!
\eeq

\medskip

Similarly to \ref{GkcalW}:
\beq\label{GkhcalW}
\encadremath{
 G^{(h)}_k(x_1,\dots,x_k) = \sum_{T\in {\cal T}_k^{(h)}}\quad {\cal W}(T)
}\eeq
where the weight ${\cal W}$ was defined in \ref{defweightW}

\subsection{Example: One-loop function, genus one}

Let us carry out explicitly the case $k=1$, $h=1$, and recover the result of \cite{eynm2m,eynm2mg1, ekk}:
In that case, \ref{loopeqgenush} reads:
\beq
\ssq{x_1}\, W^{(1)}_1(x_1)
 = {W_{2}(x_1,x_1)  + U^{(1)}_1(x_1) \over M(x_1)}
\eeq
The RHS is clearly a rational function of $x_1$, and  from \ref{Wkmcirclevanish},
we know that the LHS has poles only at the branch-points , and at $\infty$.
Introduce the Euclidean division of the polynomial $U^{(h)}_1(x_1)$ by $M(x_1)$:
\beq
U^{(1)}_1(x_1)= P^{(1)}_1(x_1)\, M(x_1) + Q^{(1)}_1(x_1)
\eeq
where $\deg P^{(1)}_1=s-2$ and $\deg Q^{(1)}_1< d-s$.

We may thus write:
\bea
\ssq{x_1}\, W^{(1)}_1(x_1)-P^{(1)}_1(x_1)
&=& \Res_{x\to x_1} {dx\over x-x_1}\,(\ssq{x}\, W^{(1)}_k(x)-P^{(1)}_1(x)) \cr
&=& \sum_{i=1}^{2s} \Res_{x\to a_i} {dx\over x_1-x}\,(\ssq{x}\, W^{(1)}_k(x)-P^{(1)}_1(x)) \cr
&=& \sum_{i=1}^{2s} \Res_{x\to a_i} {dx\over x_1-x}\,
{W_{2}(x,x)  + Q^{(1)}_1(x)\over M(x)} \cr
&=& \sum_{i=1}^{2s} \Res_{x\to a_i} {dx\over x_1-x}\, {W_{2}(x,x) \over M(x)} \cr
\eea
It clearly gives:
\bea
G^{(1)}_1(x_1)
&=& \sum_{i=1}^{2s} \Res_{x\to a_i} {G_{2}(x,x) \over y(x)dx} dS_i(x_1,x)  \cr
\eea
Diagrammatically we have:
$$
{\epsfxsize 7cm\epsffile{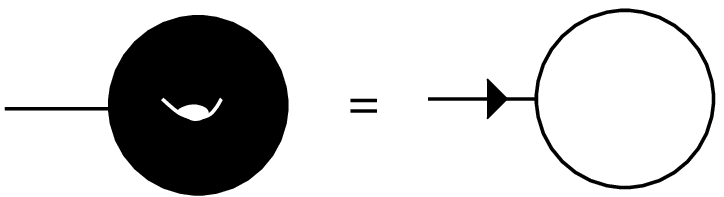}}
$$
One can check that this result is identical to  the function $W^{(1)}_1(x)$ computed in \cite{ekk}.

Similarly we have:
$$
{\epsfxsize 12cm\epsffile{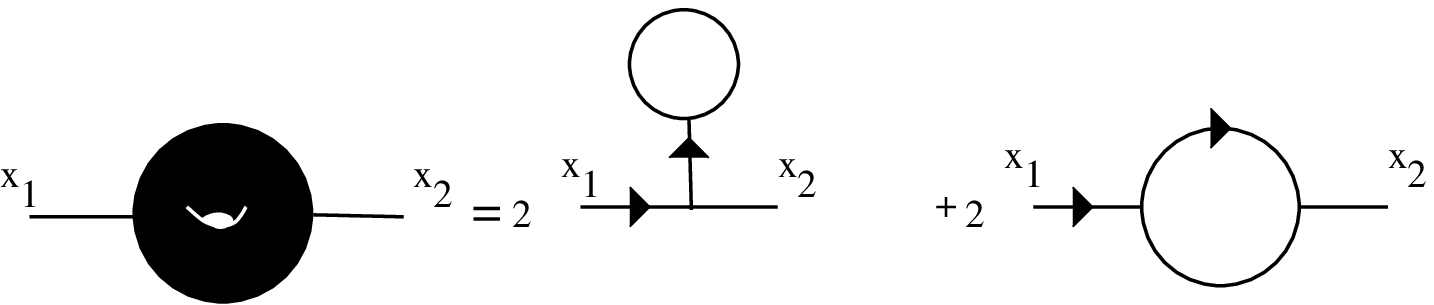}}
$$

$$
{\epsfxsize 15cm\epsffile{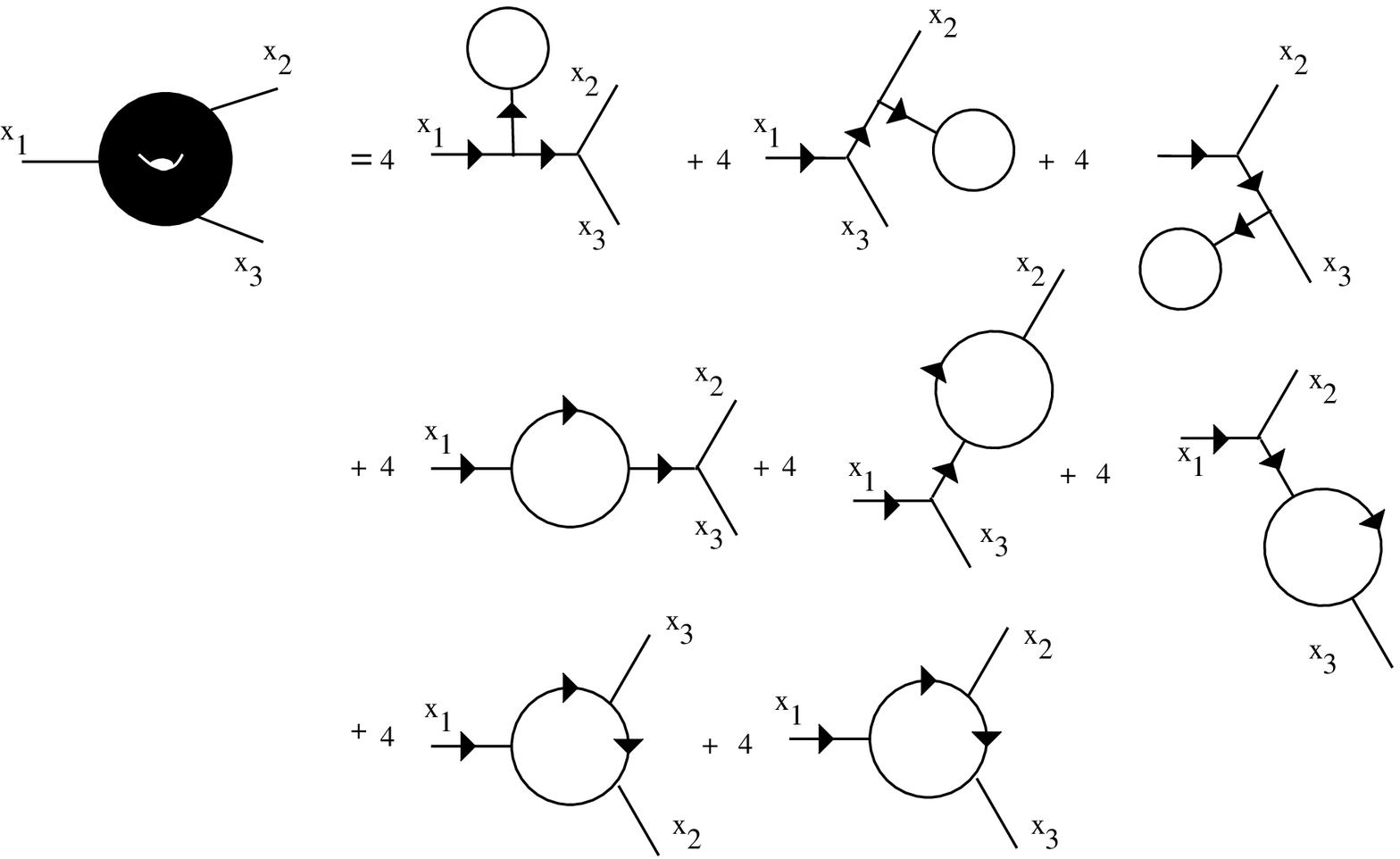}}
$$

and at genus 2 we have:
$$
{\epsfxsize 15cm\epsffile{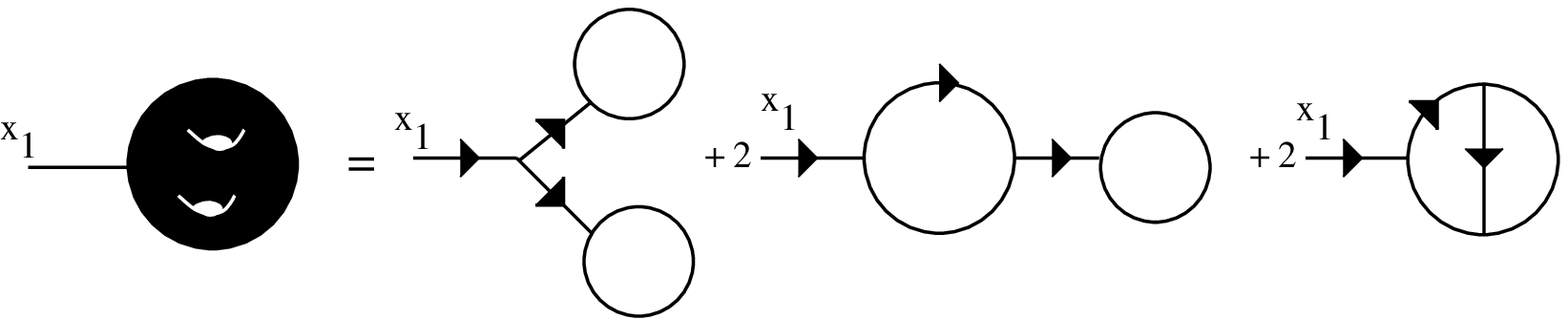}}
$$

and so on...

\newsection{Example, One-cut case s=1, i.e. genus zero curve}\label{onecut}

We write:
\beq
\sigma(x)=(x-a)(x-b)
\eeq
It is convenient to map the genus zero hyperelliptical surface into the complex plane with the rational map:
\beq
x(\l) = {a+b\over 2} + \gamma(\l+\l^{-1})
\eeq
\beq
\gamma={b-a\over 4}
\eeq
The $x$-physical sheet is sent to the exterior of the unique disc in the $\l$-plane,
and the $x$-second sheet is sent to the interior of the unique disc in the $\l$-plane.
We have:
\beq
\ssq{x(\l)} = \gamma(\l-\l^{-1})
\eeq
With this parameterization, all correlation functions are rational functions of the $\l$'s.

\subsection{Recursion relations}

All $P_k$'s are identically vanishing.
We have the formula:
\beq\label{eqrecursiong0}
\encadremath{
\begin{array}{rcl}
F_2(x_1,x_2)
& = &   2\ssq{x_2}\,{\d \over \d x_2} {\sqrt{\sigma(x_2)} \over (x_1-x_2)} = {2 x_1 x_2 - (a+b)(x_1+x_2)+2ab\over (x_1-x_2)^2} \cr
F_k(x_1,\dots,x_k)
& = & - {1\over 2}\displaystyle\mathop{\rm Res}_{a_i}
\left(   \sum_{j=1}^{k-2} \sum_{I\in K_j}
{F_{j+1}(x,x_{I}) F_{k-j}(x,x_{K-I})
\over (x-x_1)M(x)\sigma(x) } \right) dx  \,\,\,  {\rm for}\,\, k\geq 3
\end{array}
}\eeq

and for $k+h>1$, \ref{mainrecGkh} gives:
\beq\label{mainrecGkhgzero}
\encadremath{
\begin{array}{lll}
 F^{(h)}_k(x_1,x_K)
&=& 2\sum_{i=1}^{2s}\,\Res_{x\to a_i} \sum_{m=0}^{h-1}
{F_1^{(h-m)}(x) F^{(m)}_k(x,x_K)\over (x_1-x)M(x)\sigma(x)}\,dx \cr
&&+ \sum_{i=1}^{2s}\,\Res_{x\to a_i} \sum_{m=0}^{h} \sum_{j=1}^{k-2}\sum_{I\in K_j}
{F^{(m)}_{j+1}(x,x_{I}) F^{(h-m)}_{k-j}(x,x_{K-I})\over (x_1-x)M(x)\sigma(x)}\,dx  \cr
&& +\sum_{i=1}^{2s}\,\Res_{x\to a_i}  {F^{(h-1)}_{k+1}(x,x,x_K)\over (x_1-x)M(x)\sigma(x)}\,dx  \cr
\end{array}
}\eeq

\subsection{2 point function}

The 2-point function can be written as:
\beq
W_2(x_1,x_2) = -{\d \over \d x_1}{\d \over \d x_2} \ln{(\l_1-\l_2^{-1})}
= {\d \over \d x_1}{\d \over \d x_2} \ln{\left({\l_1-\l_2\over x_1-x_2}\right)}
\eeq
where
\beq
x_1 = \gamma(\l_1+\l_1^{-1})
\virg
x_2 = \gamma(\l_2+\l_2^{-1})
\eeq
or:
\bea
W_2(x_1,x_2)
& = & -{1\over 4\ssq{x_1}\ssq{x_2}}\left(
1-\left(
{\ssq{x_1}-\ssq{x_2}\over x_1-x_2}
\right)^2 \right) \cr
& = & -{1\over 2(x_1-x_2)^2}
+{2 x_1 x_2 - (a+b)(x_1+x_2)+2ab\over 4(x_1-x_2)^2 \ssq{x_1}\ssq{x_2}}
\eea

In particular we have:
\beq
F_2(a,x) = {(a-b) \over (x-a)}
\virg
F_2(b,x) = {(b-a) \over (x-b)}
\eeq

The polynomial $A(x_1,x_2)$ introduced in \ref{defQB} vanishes identically, and we have:
\beq
W_2(x,x)={(b-a)^2\over 16\,\sigma(x)^2}
\eeq
All this is well known, see for instance \cite{akemanambjorn}.

\subsection{Other correlation functions}

We just give some examples of applications of the general theory described above:

\bea
F_3(x_1,x_2,x_3)
& = & - \displaystyle\mathop{\rm Res}_{a,b}
\left(
{F_{2}(x,x_2) F_{2}(x,x_3)
\over (x-x_1)M(x)\sigma(x) } \right) dx \cr
& = & -  {F_{2}(a,x_2) F_{2}(a,x_3)
\over (a-x_1)M(a)(a-b) }  \cr
&   & -  {F_{2}(b,x_2) F_{2}(b,x_3)
\over (b-x_1)M(b)(b-a) }  \cr
& = & {b-a \over (a-x_1)(a-x_2)(a-x_3)M(a) }  \cr
&   & -{b-a \over (b-x_1)(b-x_2)(b-x_3)M(b) }  \cr
\eea
i.e.
\beq
 W_3(x_1,x_2,x_3)
 =  (b-a){
  {1 \over (a-x_1)(a-x_2)(a-x_3) M(a) }
    - {1\over  (b-x_1)(b-x_2)(b-x_3) M(b) }
 \over 8\ssq{x_1} \ssq{x_2}\ssq{x_3} }
\eeq
which is the usual of \cite{akemanambjorn}.

\bea
\ssq{x_1}\,W^{(1)}(x_1)
&=& \Res_{a,b} W_2(x,x) {dx\over 2 (x_1-x) M(x)} \cr
&=& {(b-a)^2\over 32}\Res_{a,b} {dx\over (x_1-x)M(x) \sigma(x)^2} \cr
&=& {(b-a)^2\over 32}\Res_{a} {dx\over (x-a)^2}{1\over  (x_1-x)M(x) (x-b)^2} \cr
&& +{(b-a)^2\over 32}\Res_{b} {dx\over (x-b)^2}{1\over  (x_1-x)M(x) (x-a)^2} \cr
&=& {(b-a)^2\over 32}\left.\left({1\over  (x_1-x)M(x) (x-b)^2}\right)'\right|_{x=a} \cr
&& +{(b-a)^2\over 32}\left.\left({1\over  (x_1-x)M(x) (x-a)^2}\right)'\right|_{x=b} \cr
&=& {(b-a)^2\over 32}\left({1\over M(a)}\,{-2x_1-b+3a\over (a-b)^3(x_1-a)^2}-{M'(a)\over M(a)^2}\,{1\over  (a-b)^2(x_1-a)}\right) \cr
&& +{(b-a)^2\over 32}\left({1\over M(b)}\,{-2x_1-a+3b\over (b-a)^3(x_1-b)^2}-{M'(b)\over M(b)^2}\,{1\over  (b-a)^2(x_1-b)}\right) \cr
\eea
which again agrees with \cite{ACKM} and other results in the literature.

\newsection{Conclusions and prospects}\label{concl}

In this article, we have found a $\phi^3$ Feynmann graph formulation for computing all correlations functions to all powers of $N$ in the one-hermitian matrix model.
First, it would be interesting to find out to which field theory it corresponds.
One is tempted to compare with Liouville's theory (which is not cubic) or to a fermionic theory.

We claim that this approach is more efficient for actual calculations, than the method existing previously in the literature \cite{akeman, ACKM}.
Indeed, in \cite{akeman, ACKM}, one has to construct the correlation functions recursively, by expanding them on basis functions which are themselves constructed recursively by taking derivatives with respect to the potential.
For instance, one does not get any simplification in the method of \cite{akeman, ACKM} by assuming an even potential, or by assuming a quadratic potential.
The method presented here, works for fixed potential (for instance quadratic), and does not need to construct any basis of functions.

Another important point for the method presented here, is that it is expressed in terms of geometrical fundamental objects on the spectral curve.
This is another evidence of the deep link between tau functions and complex geometry.

There are other expressions in the literature involving Residues of geometrical objects (for instance \cite{WZ, Kric, Bertolafreeenergy, Marco, dubrovin, ekk, kostov, KazMar}), namely,
only the Bergmann kernel and not the abelian differential.
However, we claim that it should be simpler to compute the residue of a function with a simple pole (the abelian differential), than the residue of a function with a double pole (Bergmann kernel).

\bigskip

Moreover, the whole procedure described here, can be applied with very small adaptations to other matrix models, in particular the 2-matrix model,
and to non-hermitian matrix models (in particular $\beta=1,2,4$ models), this work is in progress and will be available shortly \cite{feyn2mm}.
In the 2-matrix model with potentials of degree $d_1+1$ and $d_2+1$, the computation of correlation functions of the first matrix only involves
$d_2$ vertices (i.e. cubic, quartic, ..., $d_2+2$--legs--vertex), instead of only one cubic vertex equal to $1/2y(x)$ for the 1-matrix model.
This will be further explained in \cite{feyn2mm}.

\bigskip

The observable we have not computed in this article is the free energy:
\beq
\ovl{G}_0:=-{1\over N^2}\ln{Z} :=\sum_{h=0}^\infty N^{-2h}\, G_0^{(h)}
\eeq
The free energy does not appear in the loop equations. It satisfies:
\beq
{\partial G_0^{(h)}\over \partial V(x_1)} = -W_1^{(h)}(x_1)
\eeq
therefore, in order to compute the free energy, one has  to integrate with respect to the potential, i.e. one can no longer keep the potential constant.
One would reasonably make the following conjecture for $h\geq 2$:
\beq
 G_0^{(h)} \to \sum_{T\in {\cal T}_0^{(h)}}\quad {\cal W}(T)
\eeq
for example for $h=2$
$$
{\epsfxsize 12cm\epsffile{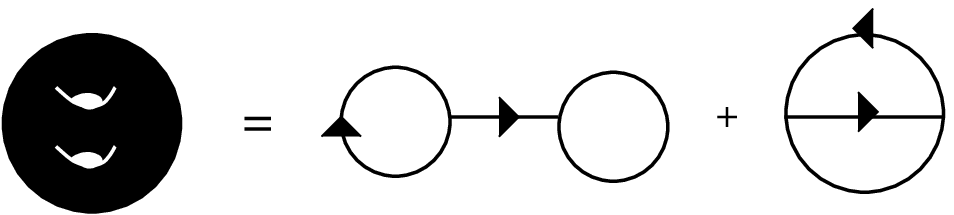}}
$$
unfortunately, these Feynmann graph don't make sense (the abelian differential diverges at coinciding points).
The conjecture is that the $ G_0^{(h)}$ are related to traces of powers of the laplacian on the spectral curve.
For instance it is known that $ G_0^{(1)}$ is related to the determinant of the laplacian \cite{ekk}.

\bigskip
{\bf \noindent Aknowledgements:}
The author wants to thank the EU network EC IHP network (HPRN-CT-1999-000161), as well as the CRM in Montreal where a large part of this research was conducted.
The author wants to thank M. Bertola, P. Di Francesco, E. Guitter, J. Harnad, I. Kostov, P. Wiegman, and A. Zabrodin for helpfull and stimulating discussions.

\setcounter{section}{0}

\appendix{Cardinal of ${\cal T}_k^{(h)}$}
\label{appNkh}

The cardinal of ${\cal T}_k$ and of ${\cal T}_k^{(h)}$ can be computed by setting ${\cal W}=1$ in \ref{GkcalW} and in \ref{GkhcalW},
and then using the recursion relations \ref{mainrecGk} and \ref{mainrecGkh}.

One thus gets for $k\geq 3$:
\beq\label{recNk}
N_2=1
\virg
N_k=\sum_{j=1}^{k-2}\,\, \pmatrix{k-1\cr j} \,\, N_{j+1}\, N_{k-j}
\eeq
writing:
\beq
r_0:=0
\virg
r_1:=1
\virg
r_k:={N_{k+1}\over k!}
\eeq
\ref{recNk} becomes for $k\geq 2$:
\beq\label{recrk}
r_{k}=\sum_{j=0}^{k}\,\,  r_j\, r_{k-j}
\eeq
We introduce the generating function:
\beq
R(x):=\sum_{k=0}^\infty r_k\, x^k
\eeq
and thus \ref{recrk} becomes:
\beq
R(x)-x=R^2(x)
\eeq
whose solution is:
\beq\label{solRx}
R(x)={1-\sqrt{1-4x}\over 2}=-{1\over 2} \,\sum_{k=1}^\infty \,\,\pmatrix{{1\over 2}\cr k}\,(-4x)^k
\eeq
which implies:
\beq
r_k =-{(-4)^k\over 2} \,\,\pmatrix{{1\over 2}\cr k}
=(-1)^{k+1}\, 2^{2k-1} {{1\over 2}\left(-{1\over 2}\right)\dots\left({3\over 2}-k\right)\over k!}
= 2^{k-1}\,{(2k-3)!!\over  k!}
= {2k-2!\over k! k-1!}
\eeq
and thus, we obtain \ref{nbtrees}
\beq
N_k=2^{k-2}\,(2k-5)!!= {2k-4!\over k-2!}
\eeq

\bigskip

For higher genus, we have for $k\geq 1$ and $h\geq 1$:
\beq\label{recNkh}
N^{(0)}_1:=0
\virg
N^{(h)}_k= N^{(h-1)}_{k+1}+\sum_{j=0}^{k-1}\sum_{m=0}^{h}\,\, \pmatrix{k-1\cr j} \,\, N^{(m)}_{j+1}\, N^{(h-m)}_{k-j}
\eeq
writing:
\beq
r^{(0)}_0:=0
\virg
r^{(h)}_k:={N^{(h)}_{k+1}\over k!}
\eeq
\ref{recNkh} becomes for $k\geq 0$, $h\geq 1$:
\beq\label{recrkh}
r^{(h)}_{k}= (k+1) r^{(h-1)}_{k+1}+\sum_{j=0}^{k}\sum_{m=0}^{h} \,\, r^{(m)}_{j}\, r^{(h-m)}_{k-j}
\eeq
We introduce the generating function:
\beq
R_h(x):=\sum_{k=0}^\infty r^{(h)}_k\, x^k
\eeq
and thus \ref{recrkh} becomes for $h\geq 0$:
\beq
R_0(x)=R(x)
\virg
R_h(x)=R_{h-1}'(x)+\sum_{m=0}^h R_m(x)\, R_{h-m}(x)
\eeq
which can also be written for $h\geq 1$:
\beq
(1-2R(x))\, R_h(x)=R_{h-1}'(x)+\sum_{m=1}^{h-1} R_m(x)\, R_{h-m}(x)
\eeq
using \ref{solRx}, it is easy to see, by induction on $h$ that for $h\geq 1$ one has:
\beq
R_h(x)=s_h\,\,\left(1-4x\right)^{-{3h-1\over 2}}
\eeq
where the coefficients $s_h$ obey for $h\geq 1$:
\beq
s_1=1
\virg
s_h=2(3h-4)s_{h-1}+\sum_{m=1}^{h-1} s_m s_{h-m}
\eeq
or, if we define $s_0:=-{1\over 2}$, it can be written for any $h\geq 1$:
\beq
0=2(3h-4)s_{h-1}+\sum_{m=0}^{h} s_m s_{h-m}
\eeq
we introduce the generating function:
\beq
S(x):=\sum_{h=0}^\infty s_h\, x^h
\eeq
it obeys:
\beq
0=S^2(x)-{1\over 4}+6x^2 S'(x)-2xS(x)
\eeq
If one writes
\beq
\xi={x^{-2/3}\over 4}
\eeq
and
\beq
S(x)=-x^{1\over 3}h(\xi)
\eeq
one has:
\beq
\xi=h^2(\xi)+h'(\xi)
\eeq
whose solution is;
\beq
h(\xi)={Ai'(\xi)\over Ai(\xi)}
={\int t\,dt\, \ee{-{t^3\over 3}+t\xi}\over \int dt\, \ee{-{t^3\over 3}+t\xi}}
=\sqrt\xi {\int t\,dt\, \ee{\xi^{3/2}(-{t^3\over 3}+t)}\over \int dt\, \ee{\xi^{3/2}(-{t^3\over 3}+t)}}
={x^{-1/3}\over 2}\, {\int t\,dt\, \ee{{1\over 8x}(-{t^3\over 3}+t)}\over \int dt\, \ee{{1\over 8x}(-{t^3\over 3}+t)}}
\eeq
and thus:
\beq
S(x)
=-{1\over 2}\, {\int t\,dt\, \ee{{1\over 8x}(-{t^3\over 3}+t)}\over \int dt\, \ee{{1\over 8x}(-{t^3\over 3}+t)}}
=-{1\over 2}\, \left( 1+ {\int t\,dt\, \ee{{1\over 8x}(-t^2-{t^3\over 3})}\over \int dt\, \ee{{1\over 8x}(-t^2-{t^3\over 3})}} \right)
\eeq
or
\beq
S(x)
=-{1\over 2}- \sqrt{x}\, {\int t\,dt\, \ee{-{t^2\over 2}}\,\ee{-{\sqrt{x}t^3\over 3}}\over \int dt\,\ee{-{t^2\over 2}}\,\ee{-{\sqrt{x}t^3\over 3}}}
\eeq
\beq
S(x)=-{1\over 2}+{\sum_{h=0}^\infty {x^{h+1}\over 3^{2h+1} \,(2h+1)!}\, \int dt \,\, t^{2(3h+2)}\, \ee{-t^2/2}
\over \sum_{h=0}^\infty {x^{h}\over 3^{2h} \,(2h)!}\, \int dt \,\, t^{2(3h)}\, \ee{-t^2/2}}
\eeq
\beq
S(x)
=-{1\over 2}+{\sum_{h=0}^\infty {x^{h+1}\, (6h+3)!!\over 3^{2h+1} \,(2h+1)!}
\over 1+\sum_{h=1}^\infty {x^{h}\, (6h-1)!!\over 3^{2h} \,(2h)!}}
=-{1\over 2}+x\,{\sum_{h=0}^\infty {x^{h}\, (6h+4)!\over 3^{2h+1}\,2^{3h+2} \,(2h+1)!\,(3h+2)!}
\over \sum_{h=0}^\infty {x^{h}\, (6h)!\over 3^{2h}\,2^{3h} \,(2h)!\,(3h)!}}
\eeq

In the end we have:
\beq
N_k^{(h)}:={\rm Card}\, {\cal T}_k^{(h)} = s_h\, (k-1)!\,4^{k-1}\,\pmatrix{{3(h-1)\over 2}+k-1\cr k-1}
\eeq

The whole function is thus:
\bea
R(x,z)&:=& \sum_k \sum_h N_k^{(h)} x^k z^h = \sum_h R_h(x) z^h \cr
&=& -z\,{\int dt\,\, t \,\, \ee{-{z^2\over 3} t^3}\,\ee{{t\over 4}(1-4x)} \over \int dt\,\,  \ee{-{z^2\over 3} t^3}\,\ee{{t\over 4}(1-4x)}}
\eea


\end{document}